\documentclass[11pt]{article}
\usepackage[authoryear]{natbib}
\usepackage{graphicx, setspace, amsmath, amssymb, subfigure, url, multirow, booktabs, verbatim, bm,threeparttable, amsthm,bbm,algorithm, color}
\usepackage{enumitem}
\usepackage[symbol]{footmisc}

\allowdisplaybreaks[3]

\def\delete#1{\iffalse #1 \fi}

\def\bse{\begin{eqnarray*}}
\def\ese{\end{eqnarray*}}
\def\bee{\begin{enumerate}}
\def\eee{\end{enumerate}}
\def\bqe{\begin{eqnarray}}
\def\eqe{\end{eqnarray}}
\def\bed{\begin{description}}
\def\eed{\end{description}}
\def\bei{\begin{itemize}}
\def\eei{\end{itemize}}

\def\pmb#1{\setbox0=\hbox{#1}%
    \kern-.025em\copy0\kern-\wd0
    \kern.05em\copy0\kern-\wd0
    \kern-.025em\raise.0433em\box0 }
\def\pmbh#1#2{\setbox0=\hbox{#1}%
    \setbox1=\hbox{#2}%
    \kern-.025em\copy0\kern-\wd0
    \kern.05em\copy1\kern-\wd0
    \kern-.025em\raise.0433em\box0 }

\def\frac#1#2{{#1\over#2}}
\def\dfrac#1#2{{\displaystyle{#1\over#2}}}

\def\boxit#1{\vbox{\hrule\hbox{\vrule\kern6pt
   \vbox{\kern6pt#1\kern6pt}\kern6pt\vrule}\hrule}}

\def\listing#1{\vskip 4mm\begin{verbatim}\input#1 \vskip 4mm}
\def\thick#1{\hbox{\rlap{$#1$}\kern0.25pt\rlap{$#1$}\kern0.25pt$#1$}}

\def\pmbh{{\pmb h}}

\renewcommand\today{\ifcase\month\or
   Jan\or Feb\or Mar\or Apr\or May\or
   Jun\or Jul\or Aug\or Sep\or Oct\or Nov\or
   Dec\fi
   \space\number\day, \number\year}

\begin{document}
\setlength{\textheight}{575pt}
\setlength{\baselineskip}{23pt}

\title{Multidimensional molecular changes-environment interaction analysis for disease outcomes}
\vspace{1cm}
\author{Yaqing Xu$^1$, Mengyun Wu$^{2*}$, Shuangge Ma$^{1*}$\\ \\
$^1$ Department of Biostatistics, Yale School of Public Health\\
$^2$ School of Statistics and Management, Shanghai University of Finance and Economics \\
Email: wu.mengyun@mail.shufe.edu.cn \\
Email: shuangge.ma@yale.edu
}

\markboth%
{Y. Xu and others}
{M-E interaction analysis}

\maketitle

\footnotetext[1]{To whom correspondence should be addressed.}

\begin{abstract}
{For the outcomes and phenotypes of complex diseases, multiple types of molecular (genetic, genomic, epigenetic, etc.) changes, environmental risk factors, and their interactions have been found to have important contributions. In each of the existing studies, only the interactions between one type of molecular changes and environmental risk factors have been analyzed. In recent biomedical studies, multidimensional profiling, under which data on multiple types of molecular changes is collected on the same subjects, is becoming popular. A myriad of recent studies have shown that collectively analyzing multiple types of molecular changes is not only biologically sensible but also leads to improved estimation and prediction. In this study, we conduct M-E interaction analysis, with M standing for multidimensional molecular changes and E standing for environmental risk factors, which can accommodate multiple types of molecular measurements and sufficiently account for their overlapping information (attributable to regulations) as well as independent information. The proposed approach is based on the penalization technique, has a solid statistical ground, and can be effectively realized. Extensive simulation shows that it outperforms multiple closely relevant alternatives. In the analysis of TCGA (The Cancer Genome Atlas) data on lung adenocarcinoma and cutaneous melanoma, sensible findings with superior stability and prediction are made.}
\end{abstract}
\noindent{\bf Keywords:} Interaction analysis, multidimensional molecular data, environmental risk factors, high-dimensional estimation

\section{Introduction}
\label{sec1}
For the outcomes and phenotypes of cancer, cardiovascular diseases, asthma, mental disorders, and other complex diseases, accumulating evidences have shown that multiple types of molecular changes, environmental risk factors, and their interactions play important roles. For example, the expression of gene IL9 is found to interact with environmental dust mite to increase severe asthma exacerbations in children \citep{Sordillo15}. In the study of lung cancer genetics, it has been suggested that smoking can act through increasing the CNV (copy number variation) of gene IGF1 to induce its oncogenesis \citep{Huang2011}. Epigenetic changes have also been investigated. For example, \cite{Teschendorff15} finds that smoking-associated DNA methylation changes in buccal cells are associated with epithelial cancers. It is observed that in each of the aforementioned and other published studies, only the interactions between a single type of molecular changes and environmental risk factors have been analyzed.

In recent biomedical studies, multidimensional profiling is becoming popular. In such studies, data on multiple types of molecular changes is collected on the same subjects. Such studies make it possible to not only more deeply understand disease biology but also construct more effective models for disease outcomes and phenotypes. A myriad of novel statistical methods have been developed. For example, \cite{wang2012ibag} proposes an integrative Bayesian analysis to identify gene expression and methylation measurements that are associated with clinical outcomes such as survival. \cite{gross2014collaborative} develops collaborative regression which applies penalization to explicitly accommodate the correlations (overlapping information) as well as independent information between gene expressions and CNVs for marker identification. \cite{zhu2016integrating} develops a linear regulatory module-based method using the sparse SVD (singular value decomposition) and penalization techniques to integrate gene expressions and their regulators for cancer outcomes. We refer to \cite{kristensen2014principles} and \cite{wu19selective} for more discussions. The aforementioned and other published studies have convincingly shown that integrating multidimensional molecular data not only is biologically sensible but also improves estimation, marker identification, and prediction. It is noted that these studies have focused on the main effects of molecular changes.

Motivated by the successes as well as limitations of the existing studies, here we conduct M-E interaction analysis, where M stands for multidimensional molecular changes and E stands for environmental risk factors. The objective is to collectively accommodate multiple types of high-dimensional molecular changes, environmental risk factors, and their interactions in modeling disease outcomes and phenotypes. This analysis is the natural next step of the integrated analysis of the main effects of multidimensional molecular data and studies that conduct the interaction analysis of a single type of molecular changes and environmental risk factors. Beyond the ``ordinary" high dimensionality and noisy nature of molecular data, the analysis faces other challenges. Specifically, multiple types of molecular measurements are interconnected, which leads to overlapping information. For example, gene expression levels are regulated by genetic and epigenetic changes. On the other hand, they can also have independent information for disease outcomes \citep{risch2008lung}. Several techniques, for example built on canonical correlation analysis \citep{Meng16} and matrix factorization \citep{zhang19}, have been developed to accommodate such overlapping and independent information. In addition, interaction analysis demands respecting the unique ``main effects, interactions'' hierarchy \citep{bien2013lasso,Wu18}, for which multiple regularization techniques have been developed.

This study has the potential to significantly expand the gene-environment interaction analysis and multidimensional molecular data analysis paradigms. The proposed approach is designed tailored to the M-E analysis and will significantly advance from the aforementioned ones. With the growing popularity of multidimensional profiling, this study can open a new venue for modeling complex diseases.

\section{Methods}
\label{sec2}

The proposed approach can accommodate multiples types/combinations of molecular measurements. Without loss of generality and to avoid confusion with terminologies, we use gene expressions and their regulators (for example, genetic and epigenetic changes) as an example in description. Such a combination has been quite popular in published studies \citep{wang2012ibag, zhu2016integrating}. Other combinations, for example proteins and gene expressions, can be analyzed in the same manner. Assume $n$ iid subjects. Denote $\boldsymbol{G}=(\boldsymbol{G}_1, \cdots, \boldsymbol{G}_{p})$ and $\boldsymbol{R}=(\boldsymbol{R}_1, \cdots, \boldsymbol{R}_{q})$ as the $n\times p$ and $n\times q$ design matrices of $p$ gene expression and $q$ regulator measurements. Denote $\boldsymbol{E}=(\boldsymbol{E}_1, \cdots, \boldsymbol{E}_M)$ as the $n\times M$ design matrix of environmental risk factors, and $\boldsymbol{Y}$ as the length $n$ vector of outcome. We first consider continuous outcomes and will discuss accommodating other types of outcomes later. Assume $\boldsymbol{Y}$ has been properly centered, and $\boldsymbol{E}, \boldsymbol{X}$, and $\boldsymbol{Z}$ have been standardized.

\subsection{M-E interaction analysis}
Our goal is to identify important M-E interactions (as well as main effects) and construct a comprehensive outcome model. Overall, the proposed approach consists of the following main steps: (i) identification of the gene expression-regulator regulatory modules, which describe the regulation relationships (overlapping information), (ii) integration of multidimensional molecular measurements within the regulatory modules, and (iii) joint modeling and estimation that respect the ``main effects, interactions'' hierarchy. The analysis flowchart is provided in Figure A1 (Appendix).

\noindent \textbf{Step I} We employ a penalized regression to estimate the gene expression-regulator regulations and then sequentially conduct biclustering to identify the regulatory modules. Consider the model
$\boldsymbol{G}=\boldsymbol{R}\boldsymbol{\Theta}+\boldsymbol{\epsilon},$
where $\boldsymbol{\epsilon}$ is the $n\times p$ matrix of random errors and $\boldsymbol{\Theta}=(\boldsymbol{\theta}_1, \cdots, \boldsymbol{\theta}_p)$ is the $q\times p$ unknown coefficient matrix. For estimating $\boldsymbol{\Theta}$, consider
\begin{equation} \label{eq:1}
\boldsymbol{\hat{\Theta}}=\arg\min_{\boldsymbol{\Theta}}\ \dfrac{1}{2}||\boldsymbol{G}-\boldsymbol{R}\boldsymbol{\Theta}||^2_F
+\lambda\sum_{j=1}^p{||\boldsymbol{\theta}_j||}_1,
\end{equation}
where $||\cdot||_F$ and $||\cdot||_1$ denote the Frobenius norm of a matrix and $L_1$ norm of a vector, and $\lambda\ge0$ is the tuning parameter.

To identify the regulatory modules, we propose conducting biclustering with $\hat{\boldsymbol{\Theta}}$. Here a regulatory module corresponds to a bicluster, which contains a small number of co-expressed gene expressions and their regulators. Specifically, for estimation, we adopt the sparse clustering technique developed in \cite{Helgeson19}, which first introduces weights for gene expressions and then maximizes the weighted between-cluster distance for regulators. The objective function is
\begin{align} \label{eq:2}
\max_{\mathcal{C}, \bar{\mathcal{C}}, \boldsymbol{w}}  &\sum_{j=1}^p w_j\left(\frac{1}{q} \sum_{l=1}^q\sum_{l'=1}^qd_{l,l',j}-\frac{1}{q_1}\sum_{l, l' \in \mathcal{C}} d_{l,l',j}-\frac{1}{q_2}\sum_{l, l' \in \bar{\mathcal{C}}} d_{l,l',j}\right),                            \\\notag
\text{subject to }
        &||\boldsymbol{w}||^2\leq 1, ||\boldsymbol{w}||_1\leq \sqrt{p}, \text{ and }w_j\geq 0  \text{ for }j=1, \cdots, p,\notag
\end{align}
where $\hat\theta_{lj}$ is the $(l,j)$th component of $\hat{\boldsymbol{\Theta}}$, $d_{l,l',j}=\left(\hat\theta_{lj}-\hat\theta_{l'j}\right)^2$ measures the distance between the $l$th and $l'$th regulators, $\mathcal{C}$ and $\bar{\mathcal{C}}$ are the disjoint index sets of regulator clusters, $q_1=|\mathcal{C}|$ and $q_2=|\bar{\mathcal{C}}|$ are the cardinalities of $\mathcal{C}$ and $\bar{\mathcal{C}}$ with $q_1< q_2$ and $q_1+q_2=q$, and $\boldsymbol{w}=(w_1, \cdots, w_p)'$ is the weight vector for gene expressions, with a larger weight indicating higher importance for clustering. With the constraints for $\boldsymbol{w}$, each $w_j$ has a nonzero value between 0 and 1. With the estimated weight $\hat{\boldsymbol{w}}$, a two-sample permutation-based Kolmogorov-Smirnov test is conducted to test the significance of the difference between two clusters and select the gene expression set $\mathcal{D}$ with significantly large weights. This process leads to one regulatory module $\{\mathcal{C},\mathcal{D}\}$ with regulators in $\mathcal{C}$ and gene expressions in $\mathcal{D}$. To obtain subsequent modules, we update $\hat{\boldsymbol{\Theta}}$ by subtracting the module just identified and repeat the above procedure. This process is iterated until the Kolmogorov-Smirnov test fails to reject the null hypothesis of no clusters. With the sparsity of $\hat{\boldsymbol{\Theta}}$, it is expected that only a subset of gene expressions and regulators can form modules. Suppose that there are $S$ identified modules $\left\{\mathcal{C}_1,\mathcal{D}_1\right\},\cdots, \left\{\mathcal{C}_S,\mathcal{D}_S\right\}$.

\noindent\underline{Rationale}
Linear regression is used to describe the regulations between two types of molecular measurements. Multiple published studies \citep{shi2015deciphering,zhu2016integrating} have shown that it is a sensible choice, especially considering the high dimensionality. One gene expression is regulated by only a few regulators, and one regulator affects the expressions of only a few genes. As such, $\boldsymbol{\Theta}$ is assumed to be sparse, and the Lasso penalization is applied for estimation and identification of important regulations.

The concept of regulatory module has been developed in \cite{zhu2016integrating} and other studies. A regulatory module consists of a small number of gene expressions and regulators that behave in a coordinated manner. The construction in \cite{zhu2016integrating}, which is based on sparse SVD, limits each regulatory module to have rank one. Here we lift this inconvenient constraint via biclustering. By construction, each bicluster (regulatory module) consists of gene expressions and regulators sharing similar patterns in $\boldsymbol{\Theta}$. We adopt the sparse biclustering method developed in \cite{Helgeson19} because of its favorable numerical performance. Note that here we cluster regulators into two disjoint groups with weighted gene expressions. It is also possible to reverse the roles of gene expressions and regulators, and this leads to similar clustering results in our numerical investigations. With the sequential cluster construction strategy, different regulatory modules may have overlaps. This is desirable as one gene/regulator can participate in multiple biological processes.

\noindent\textbf{Step II} We integrates information within each regulatory module $\left\{\mathcal{C}_s,\mathcal{D}_s\right\}, s=1,\cdots, S$, using the PCA (principle component analysis) technique. Given a matrix $\boldsymbol{A}$ and index set $\mathcal{I}$, denote $\boldsymbol{A}_{\mathcal{I}}$ as the columns of $\boldsymbol{A}$  indexed by $\mathcal{I}$. For the $s$th module, we apply PCA to the stacked matrix $(\boldsymbol{G}_{\mathcal{D}_{s}}, \boldsymbol{R}_{\mathcal{C}_{s}})$ and select the top PCs with the cumulative variance contribution rate $\geq 80\%$. Denote the resulted matrix composed of the $p_s$ PCs as $\boldsymbol{X}_{s}=\left(\boldsymbol{X}_{s,1}, \cdots, \boldsymbol{X}_{s, p_{s}}\right) $. In addition, for gene expressions and regulators not involved in any identified modules, we collect and combine them as $\boldsymbol{Z}=(\boldsymbol{Z}_1, \cdots, \boldsymbol{Z}_{p_z})=(\boldsymbol{G}_{\mathcal{D}^c},
\boldsymbol{R}_{\mathcal{C}^c})$, where $\mathcal{D}^c=\{j\in\{1,\cdots,p\}: j\notin \mathcal{D}_s, s=1,\cdots,S\}$ and $\mathcal{C}^c=\{j\in\{1,\cdots,q\}: j\notin \mathcal{C}_s, s=1,\cdots,S\}$. $\boldsymbol{X}=(\boldsymbol{X}_{1},\cdots, \boldsymbol{X}_{S})$ and $\boldsymbol{Z}$ form input for downstream analysis.

\noindent\underline{Rationale}
The previous step of analysis does not directly limit the sizes of the modules. Thus, it is possible some modules have moderate to large sizes. In addition, with regulations, measurements within the same modules often times have strong correlations. To reduce dimensionality, remove collinearity, and simplify computation, we apply PCA, which can be replaced by other dimension reduction techniques. Overall, the input for the next step consists of the PCs (representing overlapping information) and the gene expressions and regulators that do not form patterns (representing independent information).

\noindent\textbf{Step III} Here we conduct interaction analysis, that respects the ``main effects, interactions'' hierarchy \citep{bien2013lasso}. For the continuous outcome, consider the regression model
\begin{eqnarray} \label{eq:joint1}
\nonumber \boldsymbol{Y}&=&\boldsymbol{E}\boldsymbol{\alpha}+
\sum_{s=1}^S\boldsymbol{X}_s\boldsymbol{\beta}_s+
\boldsymbol{Z}\boldsymbol{\gamma}
+\sum\limits_{m=1}^M\sum_{s=1}^S (\boldsymbol{E}'_m \odot \boldsymbol{X}'_s)'(\boldsymbol{\beta}_s \ast \boldsymbol{\eta}_{sm})+\sum\limits_{m=1}^M (\boldsymbol{E}'_m \odot\boldsymbol{Z}')(\boldsymbol{\gamma}\ast \boldsymbol{\tau}_{m})+\boldsymbol{\xi},\\
&=&g(\boldsymbol{X}, \boldsymbol{Z}, \boldsymbol{E})+\boldsymbol{\xi}.
\end{eqnarray}
Here $\boldsymbol{\alpha}=(\alpha_1,\cdots,\alpha_M)'$, $\boldsymbol{\beta}=(\boldsymbol{\beta}'_1,\cdots,\boldsymbol{\beta}'_S)$, and $\boldsymbol{\gamma}=(\gamma_1,\cdots,\gamma_{p_z})'$ correspond to the main effects of the environmental factors, regulatory modules, and  individual molecular measurements (that do note belong to any module), respectively. For the $m$th environmental factor, $\boldsymbol{\beta}_s\ast\boldsymbol{\eta}_{sm}$ and $\boldsymbol{\gamma}\ast\boldsymbol{\tau}_{m}$ correspond to the interactions with the $s$th regulatory module and all individual molecular measurements, respectively, with $\ast$ being the component-wise product. $\odot$ is the ``matching column-wise'' Khatri-Rao product.  $\boldsymbol{\xi}$ is the random error vector.
Here, following the literature \citep{choi2010variable}, we use the products $E_{im} X_{ij}$ and $E_{im} Z_{ij}$ to describe the interactions for the $i$th subject. To accommodate the hierarchical structure of interaction analysis, the interaction effects $\beta_{sj}\eta_{smj}$ and $\gamma_{j}\tau_{mj}$ are decomposed into two components, the first for the corresponding main effects ($\beta_{sj}$ and $\gamma_{j}$) and the other for the interaction-specific effects ($\eta_{smj}$ and $\tau_{mj}$).

For the estimation and identification of important interactions (and main effects), we propose the penalized objective function
\begin{align}\notag
Q(\boldsymbol{\Phi})= &\ \frac{1}{2}||\boldsymbol{Y}-g(\boldsymbol{X}, \boldsymbol{Z}, \boldsymbol{E})||_2^2  \\
 &+\lambda_1\sum_{s=1}^S\sqrt{p_s}\left(||\boldsymbol{\beta}_s||_2
 +\sum_{m=1}^M||\boldsymbol{\eta}_{sm}||_2\right)+\lambda_2 \left(||\boldsymbol{\gamma}||_1+\sum_{m=1}^M
 ||\boldsymbol{\tau}_{m}||_1\right),   \label{eq:obj1}
\end{align}
where $\boldsymbol{\Phi}=(\boldsymbol{\alpha}', \boldsymbol{\beta}'_1, \cdots, \boldsymbol{\beta}'_S, \boldsymbol{\gamma}', \boldsymbol{\eta}'_{11}, \cdots, \boldsymbol{\eta}'_{MS}, \boldsymbol{\tau}'_{1}, \cdots, \boldsymbol{\tau}'_{M})'$, $||\cdot||_2$ is the $L_2$ norm of a vector, and $\lambda_1, \lambda_2 \ge 0$ are tuning parameters.
Gene expressions and regulators that are involved in modules with nonzero estimated
$\boldsymbol{\beta}_s$ and $\boldsymbol{\beta}_s\ast\boldsymbol{\eta}_{sm}$ are identified as having important main effects and M-E interactions, respectively. In addition, for individual molecular measurements, the nonzero components of $\boldsymbol{\gamma}$ and $\boldsymbol{\gamma}\ast\boldsymbol{\tau}_{m}$ correspond to important main effects and interactions, respectively.

\noindent\underline{Rationale}
A joint model is developed to accommodate all molecular and environmental effects and their interactions. As to be described below, the linear regression model can be replaced by other models. For estimation and selection, we adopt penalization, which has been the choice of quite a few recent interaction studies \citep{bien2013lasso,Wu18}. For many datasets including those analyzed in this article, the environmental factors are pre-selected based on existing knowledge and usually considered as important, so that their coefficients are not subject to penalized selection. As such, the ``main effects, interactions'' hierarchy postulates that an identified interaction corresponds to an identified main molecular effect. To achieve this, we decompose the interaction effects into two components and have that $\beta_{sj}\eta_{smj}\neq 0$ only if $\beta_{sj}\neq 0$ and $\gamma_{j}\tau_{mj}\neq 0$ only if $\gamma_{j}\neq 0$ \citep{choi2010variable}. In (\ref{eq:obj1}), we employ group Lasso for regulatory modules (where PCs corresponding to the same module form a group) and Lasso for individual molecular measurements to identify M-E interactions and main effects. Here, all PCs corresponding to the same module are in or out simultaneously, which is motivated by the coordinated nature of the molecular measurements in the same module.

\noindent\textbf{Accommodating other types of outcomes} With a different type of outcome variable, the lack-of-fit in (\ref{eq:obj1}) can be replaced by the negative log-likelihood function or an estimating equation-based measure. As an example, consider survival data, which is analyzed below. Denote $\boldsymbol{T}$ as the length $n$ vector of survival times. Consider the AFT (accelerate failure time) model $\log(\boldsymbol{T})=g(\boldsymbol{X}, \boldsymbol{Z}, \boldsymbol{E})+\boldsymbol{\xi}$, where notations have similar implications as above. Denote $\boldsymbol{C}$ as the length $n$ vector of censoring times, then we observe $\boldsymbol{Y}=\log(\min(\boldsymbol{T}, \boldsymbol{C}))$ and $\boldsymbol{\delta}=I(\boldsymbol{T}\le \boldsymbol{C})$ with $I(\cdot)$ being the indicator function. Assume that data has been sorted according to the observed times from the smallest to the largest. Compute the Kaplan-Meier weights: $\rho_1=\frac{\delta_1}{n}$, $\rho_i=\frac{\delta_i}{n-i+1}\prod\limits_{i'=1}^{i-1}(\frac{n-i'}{n-i'+1})^{\delta_{i'}}, i=2, \dots, n$. Then, we have the weighted penalized objective function
\begin{equation*}
\frac{1}{2}||\sqrt{\bm\rho}\ast(\boldsymbol{Y}-g(\boldsymbol{X}, \boldsymbol{Z}, \boldsymbol{E}))||_2^2 +\lambda_1\sum_{s=1}^S\sqrt{p_s}\left(||\bm{\beta}_s||_2+\sum_{m=1}^M||\bm{\eta}_{sm}||_2\right)+\lambda_2 \left(||\boldsymbol{\gamma}||_1+\sum_{m=1}^M||\boldsymbol{\tau}_{m}||_1\right). 
\end{equation*}

\subsection{Computation}
The detailed computational algorithms for Steps I and III are provided in Algorithms 1 and 2 (Appendix), respectively. Step II can be realized using existing algorithms and R function \texttt{prcomp}. In computation, effort has been made to take advantage of the existing algorithms and software. When not possible, optimization has been based on the CD (coordinate descent) techniques. In the literature, convergence properties of the CD and other techniques used in computation have been well established. Convergence is observed in all of our numerical studies. The two tuning parameters in \eqref{eq:obj1} are selected using the extended Bayesian information criterion \citep{chen2008extended}. The proposed algorithm is computationally feasible. For example, under a standard laptop configuration, it takes less than five minutes for a simulated dataset with 250 subjects, 500 gene expression measurements, and 500 regulator measurements. We have developed R code implementing the proposed approach and made it publicly available at \url{https://github.com/shuanggema/omics_interaction}.

\subsection{Heuristic theoretical justifications}

Consider the scenario where the number of molecular factors (gene expressions and their regulators) increases and the number of environmental factors is finite as the sample size increases. There are several key estimation procedures and conditions.
First, in the step of identifying regulatory modules, the consistency of Lasso estimator $\bm{\hat{\Theta}}$ is needed.
For each gene expression, with probability at least $1-\frac{2}{\sqrt{\pi}}qu_n^{-1}e^{-u_n^2/2}$, $\bm{\theta_j}$ can satisfy the weak oracle property, under mild regularity conditions on the design matrix $\bm{R}$, signal strengths, Gaussian random error, and $q=o(u_n e^{u_n^2/2})$. Here, the order of $u_n$ can be $o(n^{a})$ with $a\in (0,\frac{1}{2}]$, leading to $\log(q)=o(n^{2a})$. Thus, with a total of $p$ gene expressions, to ensure the overall consistency of $\bm{\hat{\Theta}}$, it is required that $1-\frac{2}{\sqrt{\pi}}q p u_n^{-1}e^{-u_n^2/2}\rightarrow 1$ with the Bonferroni approach. Assume that $p$ and $q$ are of the same order, then we have $\log(q)=\log(p)=o(n^{a})$.
Second, the adopted biclustering strategy is an ``upgrade'' of the sparse K-means clustering with an $L_ {2}/L_1$ penalty. For the sparse K-means with an $L_ {\infty}/L_0 $ penalty, it has been shown in \cite{Chang14} that under certain regularity conditions, the estimated weight $\bm{w}$ has feature selection consistency.
Consistency under an $L_ {2}/L_1$ penalty is expected to hold with revised norm assumptions, which will lead to consistency of the estimated gene expression clusters. Consistency of the estimated cluster centres of K-means has been well established in \cite{Pollard81}, which can support the consistency of the estimated regulator clusters. Combining such results is expected to lead to the consistency of biclustering. Third, for each regulatory module, PCA is conducted to extract integrated information. With the ratio $n/(|\mathcal{C}_s|+|\mathcal{D}_s|)\rightarrow 0$, \cite{Jung09} shows that if the first few eigenvalues are large enough compared to the others, then the corresponding estimated PC directions are consistent or converge to the appropriate subspace (subspace consistency). Finally, for estimators in interaction analysis with hierarchy, consistency has been established in \cite{choi2010variable} and \cite{Wu18}. As shown in \cite{Wu18}, under mild regularity conditions on the design matrix, smallest signal, and tuning parameters, the estimator has consistency properties, where the dimensionality $p+q$ is allowed to grow up exponentially faster than the sample size.

\section{Simulation}
\label{sec3}
We set $p=q=500$, $M=5$, and $n=250$, and generate environmental factors from independent standard normal distributions. In addition, (a) we consider two settings for $\boldsymbol{\Theta}$ to represent different regulation patterns. The first $(\boldsymbol{\Theta}_1)$ contains $15$ regulatory modules with one overlapping. The corresponding elements are independently generated from normal distributions with mean ranging from $-0.7$ to $1.5$ and standard deviation $0.1$, covering different levels and directions of regulations on average. Each regulatory module contains $12.3$ gene expressions and $16.6$ regulators. The rest elements of $\boldsymbol{\Theta}_1$ are zero. The second $(\boldsymbol{\Theta}_2)$ contains $20$ nonzero regulatory modules with one overlapping, and the nonzero values are generated similarly as $\boldsymbol{\Theta}_1$. Those modules consist of $6.0$ gene expressions and $8.1$ regulators on average. Compared to $\boldsymbol{\Theta}_1$, $\boldsymbol{\Theta}_2$ contains more modules with smaller sizes, representing a different type of regulations. (b) The values of regulators $\boldsymbol{R}$ involved in each regulatory module are generated from a multivariate normal distribution with marginal means 0 and variances 1. We consider three correlation structures. The first (R1) is an auto-regressive structure where the correlation between the $j$th and $l$th variables is $(-0.5)^{|j-l|}$. The second (R2) is a banded structure where the correlation between the $j$th and $l$th variables is $-0.5$ if $|j-l|=1$ and $0$ otherwise. The third (R3) has a structure where the correlation between the $j$th and $l$th variables is $(-1)^{|j-l|}/(|\mathcal{C}_s|+|\mathcal{D}_s|)$. Among them, R1 and R2 are ``diagonally dominant'', while R3 has all correlations at the same level. The individual regulators that are not involved in any regulatory modules are independently generated from the standard normal distribution. As such, regulators in different modules are independent of each other and also independent of the individual regulators. (c) Gene expression measurements are generated by $\boldsymbol{G}=\boldsymbol{R}\boldsymbol{\Theta}+\boldsymbol{\epsilon}$, where the elements of $\boldsymbol{\epsilon}$ follow independent standard normal distributions. (d) Given $\boldsymbol{G}$, $\boldsymbol{R}$, and $\boldsymbol{\Theta}$, generate the integrated information $\boldsymbol{X}_s$ for each module using the top PCs and $\boldsymbol{Z}$ for the individual molecular units. (e) With $\boldsymbol{X}_s, s=1,\cdots, S$ and $\boldsymbol{Z}$,  consider the continuous response under model (\ref{eq:joint1}). Two types of nonzero coefficient settings are considered, leading to a total of $100$ (P1) and $70$ (P2) important main molecular effects and M-E interactions, respectively. These nonzero coefficients are generated uniformly from $(0.5, 0.8)$ (B1) or $(0.8, 1.2)$ (B2), representing two signal levels, with the ``main effects, interactions'' hierarchical structure satisfied. The molecular factors with important effects include gene expressions and regulators involved in the regulatory modules as well as individual molecular measurements. Additional information is provided in the Appendix. Random errors $\boldsymbol{\xi}$ are generated from independent standard normal distributions. 

To better appreciate operating characteristics of the proposed module detection procedure, we simulate one dataset under setting $\boldsymbol{\Theta}_1$ and correlation structure R1. We present the true regulation relationships between gene expressions and regulators in Figure 1, together with their estimated values and identified regulatory modules. We observe that with moderate associations between small sets of molecular measurements, the estimated $\hat{\boldsymbol{\Theta}}$ based on Lasso closely reflects the true regulation relationships. Furthermore, biclustering is able to properly identify the regulatory modules based on the estimated regulations.

To be more informative, besides the proposed approach, we also consider the following alternatives which have closely related frameworks. Comparing with these alternatives can directly establish the necessity of the considerations on gene expression-regulator regulations, correlations within regulatory modules, and hierarchical interactions. Specifically, Alt.1 excludes Step II of integration and builds the hierarchical interaction model using gene expressions and regulators directly combined as groups based on the identified regulatory modules. Alt.2 excludes the decomposition of interaction coefficients in Step III, and so the ``main effects, interactions'' hierarchical structure may be violated. Alt.3 builds the hierarchical joint model directly using the original stacked gene expression and regulator measurements without accounting for the regulations. Alt.4 incorporates the original stacked gene expression and regulator measurements directly in the interaction model. It ignores the regulation relationships and interaction hierarchy. For evaluation, we consider the numbers of true positives (TP) and false positives (FP) for main effects and interactions together.

For each scenario, 200 replicates are simulated. Summary results under settings P1 and P2 are presented in Tables \ref{tab:t1} and \ref{tab:t2} respectively. We observe that the proposed approach achieves better or comparable performance in identification accuracy. For example in Table \ref{tab:t1} with weak effects (B1), regulation pattern $\boldsymbol{\Theta}_1$, and correlation structure R1, the proposed approach selects on average $95.94$ true positives, compared to $71.90$ (Alt.1), $65.20$ (Alt.2), $23.15$ (Alt.3), and $16.80$ (Alt.4). When there are more correlated molecular measurements, the proposed approach remains superior in identification. For instance in Table \ref{tab:t1} with weak effects (B1), regulation pattern $\boldsymbol{\Theta}_1$, and correlation structure R3, the proposed approach selects on average $99.70$ true positives with $8.50$ false positives. In comparison, Alt.1, Alt.2, Alt.3, and Alt.4 select fewer true positives and more false positives with (TP,FP)=(83.68,12.26), (95.90,54.75), (27.30,14.70), and (20.75,136.65), respectively. With a higher signal level under setting B2, all approaches behave better, while with more regulation modules under setting $\boldsymbol{\Theta}_2$, performance of all approaches decays. Under both settings, the proposed approach still has advantage. It is observed that Alt.1 generally achieves the second best identification performance, and under some scenarios it is competitive in true positive identification compared to the proposed approach, at the cost of larger numbers of false positives. This is because the integration procedure of the proposed approach that uses PCs for the joint interaction model can effectively remove collinearity and reduce false discovery. The proposed approach performs better than Alt.2, suggesting that the hierarchical interaction modeling can lead to more accurate identification. The superior performance of the proposed approach over Alt.3 and Alt.4 provides a direct support to the integrated analysis strategy that accommodating the regulations among multidimensional molecular data in interaction analysis substantially improves identification performance.

\section{Data analysis}
\label{sec4}
TCGA is one of the largest data resources with multidimensional profiling. TCGA data have been analyzed in interaction analysis with one type of molecular measurements as well as integrated modeling with the main effects of multiple types of molecular measurements. This study is the first to conduct the integrated M-E interaction analysis. We analyze data on lung adenocarcinoma (LUAD) and cutaneous melanoma (SKCM). Data are downloaded from TCGA Provisional using the R package {\it{cgdsr}}.

\subsection{Analysis of LUAD data}

The response of interest is the reference value for the pre-bronchodilator forced expiratory volume in one second in percent (FEV1). It is an important biomarker for lung capacity, with a lower value suggesting the potentially functional disorder of the lung, and has been shown to be a powerful marker for future morbidity and mortality \citep{young2007forced}. It is continuously distributed and ranges from 1.95 to 156 with mean 80.58 and standard deviation 23.55. We focus on the primary tumor samples of the Whites. For environmental risk factors, we consider age, American Joint Committee on Cancer (AJCC) tumor pathologic stage (Stage), tobacco smoking history indicator (Smoking), and gender, which have been extensively investigated in the literature. We analyze mRNA gene expression measurements which were collected using the Illumina HiSeq 2000 RNA Sequencing Version 2 analysis platform. For regulators, we include CNV measurements that were collected using the Genome-Wide Human SNP Array 6.0 platform and DNA methylation measurements that were collected using the Illumina Infinium HumanMethylation450 platform. A total of 18,345 gene expression, 23,321 CNV, and 15,288 methylation measurements are available. In principle, the proposed approach can be directly applied. However, considering that only a small number of molecular measurements are potentially associated with the outcome and the analysis may be unstable with the high dimensionality and small sample size, we conduct a prescreening. Specifically, we select the top 1,000 molecular measurements with the smallest p-values using marginal regression. This leads to 164 subjects with 467 gene expression and 533 regulator (316 CNV and 217 methelaytion) measurements for downstream analysis.

The proposed analysis identifies 20 regulatory modules in Step I, and each module on average contains $11.70$ gene expression and $7.35$ regulator measurements. The graphical presentation of the modules is provided in Figure A2 (Appendix), where some overlappings between modules are observed. In interaction analysis, the proposed approach identifies 62 main molecular effects and 29 M-E interactions, among which 50 main effects and 27 interactions belong to six regulatory modules. The identified main effects consist of 41 gene expression, 9 CNV, and 12 methylation measurements, and the identified interactions consist of 20 with gene expressions and 9 with methylations. Detailed estimation results are presented in Table \ref{tab:t3}, where a ``group" corresponds to a module or an individual measurement. Literature search suggests that the findings are biologically sensible. For example, Stage and Smoking are shown to be negatively associated with FEV1, which has also been suggested in previous studies. Gene AFF3 is identified along with its interactions with Smoking and gender. A decreased methylation of gene AFF3 in non-small cell lung tumors has been found as one of the key epigenetic changes associated with lung cancer development. Gene PWRN1 has been reported to be involved in the process of spermatogenesis, and its expression level has been shown to be related to tumor size in lung cancer patients. Gene CACNG3 has been identified as an oncogene from a pan-cancer study with somatic mutation data, suggesting its potentially important role for lung adenocarcinoma. Gene PRH1 has been identified as one of the candidate exosomal protein biomarkers for the detection of lung cancer using human saliva and serum. In addition, published studies have shown that gene CACNG6 is significantly upregulated in lung squamous cell carcinoma compared to normal lung tissues. Gene PABPC5 has been found to be hypermethylated among early-stage non-small cell lung cancer patients compared to controls. Gene MAP4K4 has been demonstrated to be frequently overexpressed in many types of human cancers, relating to transformation, invasiveness, adhesion, and cell migration. Patients with lung adenocarcinoma and high MAP4K4 expressions have been found to have a shorter overall survival. The lower expression levels of gene DRD3 have been found among patients with non-small cell lung cancer.

We take a closer look at the functional and biological connections of genes involved in each identified regulatory module. Specifically, the gene ontology (GO) enrichment analysis is conducted using DAVID version 6.8 \citep{Sherman09}. It is observed that the identified modules are biologically meaningful with certain significantly enriched GO terms. For example, in regulatory module \#1, genes CACNG6 and RYR3 are enriched with calcium channel activity (GO:0005262, p-value= 0.0042) and calcium ion transport (GO:0006816, p-value=0.0072). Biological studies have found calcium controls cell death and proliferation that are relevant to tumorigenesis, and up or down regulations of specific calcium channels and pumps are associated with cancers. As another example, genes ATP8A2 and DGUOK in regulatory module \#20 are enriched with purine nucleoside triphosphate (GO:0009144, p-value=0.0053) and purine nucleoside metabolic process (GO:0006163, p-value=0.008), suggesting the functional and biological connections within the identified module.

Analysis is also conducted using the alternative approaches. In Table A1 (Appendix), we provide the comparison results, including the numbers of identified main effects and interactions, and numbers of overlapping and RV coefficients between the identifications using different approaches. The RV coefficient measures the common information of two data matrices. It lies between 0 and 1, and a larger value indicates a higher degree of overlapping. We observe that different approaches select significantly different sets of main effects and interactions, with moderate overlapping as measured by the RV coefficients. In practical data analysis, it is difficult to objectively evaluate identification performance. To provide an indirect support, we evaluate prediction performance and selection stability. Specifically, for prediction evaluation, we consider the prediction mean squared error (PMSE) based on 200 random resamplings (9/10 training and 1/10 testing samples). The proposed approach demonstrates competitive performance with the average PMSE$=1.02$, compared to $1.25$ (Alt.1), $1.16$ (Alt.2), $1.05$ (Alt.3), and $1.02$ (Alt.4). We also assess selection stability using the observed occurrence index (OOI) \citep{huang2006regularized}. For each identified main effect (interaction), OOI computes its selection frequency in the 200 resamplings, and a larger value suggests higher stability. The proposed approach is observed to have much satisfactory stability with the average OOI value being $0.77$, compared to $0.53$ (Alt.1), $0.45$ (Alt.2), $0.26$ (Alt.3), and $0.21$ (Alt.4).

\subsection{Analysis of SKCM data}
The response of interest is overall survival, which is subject to censoring. We focus on the primary tumor samples of the Whites. We consider age, AJCC tumor pathologic stage (Stage), gender, and Clark level at diagnosis (Clark), all of which have been suggested as associated with melanoma in the literature. A total of 18,925 gene expression, 23,287 CNV, and 15,616 methylation measurements are available. With the same prescreening as in the previous analysis, the data used for downstream analysis contains 314 gene expression and 686 regulator (397 CNV and 289 methylation) measurements on 231 subjects, of which 139 died during follow-up. The observed times range from 2.04 to 357.10 months with median 56.31.

The proposed analysis identifies 17 regulatory modules, which contain on average $7.60$ gene expressions and $6.45$ regulators. The graphical presentation is provided in Figure A2 (Appendix). The AFT model is assumed for modeling survival. A total of 28 main effects and 12 interactions are selected by the proposed approach, among which 14 main effects belong to one identified regulatory module and the remaining are related to the individual molecular units. The identified main effects consist of 15 gene expression and 13 methylation measurements, and the identified interactions consist of 9 with gene expressions and 3 with methylations. The estimated coefficients are presented in Table \ref{tab:t6}. Examining the estimated coefficients suggests that melanoma patients with higher levels of age, Stage, and Clark have a shorter survival. Findings on the molecular variables are also sensible. For instance, gene IMP3 has been found to be associated with cell proliferation and  considered as an oncofetal protein-related gene. Its expression level has been used as a diagnostic and prognostic marker from surgical pathology in malignant melanoma. TBC1D7 is one of the down-regulated genes that are potentially causal for the induction of loss of proliferative capacity and terminal differentiation in human melanoma cells. The lack of gene A2M expression provides a growth advantage to melanoma cells by interfering with effective antigen presentation. IL24 is a novel tumor suppressor gene with tumor-apoptotic and immune-activating properties, and one of several genes that are upregulated during terminal differentiation of melanoma cells. The high expression level of gene ZDHHC4 has been observed in NRAS mutant melanoma cell lines. Published studies have also found a statistically significant overexpression of gene BRF2 in cutaneous melanoma compared to normal skin, and suggested it as a potential marker for patients at risk for metastasis. Gene RBP2 has been shown to directly regulate gene transcription in a reporter assay system as a transcriptional regulator with a tumor suppressive potential in melanoma cells. For the identified module, we further conduct the GO enrichment analysis. It is observed that the involved genes share common GO terms. For example, genes A2M, ENOX1, and IL24 are enriched with extracellular space (GO:0005615, p-value=0.0097), for which published studies have suggested that extracellular vesicles released to extracellular space are correlated with genetic tumor progression in human cancer.

We conduct analysis using the alternatives and summarize the comparison results in Table A1 (Appendix). Similar patterns as in the previous analysis are observed, where different approaches have small numbers of overlapping identifications and moderate RV coefficients. We also conduct the prediction and selection stability evaluation. With the censored survival response, we adopt the C statistic to measure prediction accuracy \citep{uno2011c}. A larger value of C statistic indicates better prediction. The proposed approach has an average C statistic $0.60$, compared to $0.57$ (Alt.1), $0.48$ (Alt.2), $0.47$ (Alt.3), and $0.59$ (Alt.4). In addition, it has superior selection stability with an average OOI of $0.74$, compared to $0.56$ (Alt.1), $0.50$ (Alt.2), $0.38$ (Alt.3), and $0.26$ (Alt.4). These results provide a strong support to the proposed M-E interaction analysis.

\section{Discussion}
\label{sec5}
Modeling the outcomes and phenotypes of cancer and other complex diseases is an ``old" but still widely open problem. In this study, we have developed the M-E interaction analysis, which is the natural next step of the existing literature. In particular, it is built on but advances from the existing gene-environment interaction analysis by incorporating multiple types of molecular measurements (which have overlapping but more importantly independent information in a single analysis). It also advances from the existing multidimensional molecular data analysis by incorporating interactions and respecting the hierarchical structure. The proposed approach has sound biological and statistical basis. Its working characteristics are carefully examined, and simulation and data analysis have demonstrated its satisfactory performance.

It remains an open question how to best accommodate multidimensional molecular data in modeling. The proposed analysis Step I has been motivated by \cite{wang2012ibag},  \cite{zhu2016integrating}, and several other studies. Similar to the literature, linear modeling and regularized estimation have been applied for estimating the regulations. Different from the literature, biclustering has been conducted to identify local regulations, where a small number of co-expressed genes are regulated by a small number of regulators in a coordinated manner. It advances from \cite{zhu2016integrating} and others by relaxing the rank-one constraint. The Step II of dimension reduction can be conducted by other techniques such as partial least squares, can effectively reduce dimensionality and remove collinearity, and has been shown as effective in numerical study. There are alternative techniques for interaction analysis in Step III. We have chosen penalization for the consistency of analysis framework. It will be of interest to extend by adopting other estimation/selection techniques. We have used gene expressions and their regulators for description. The proposed approach can be directly applied to other and potentially more complex data structures, thus enjoying broad applicability.

%


\section*{Acknowledgments}

This work was supported by the National Institutes of Health [CA204120, CA241699, CA216017]; National Science Foundation [1916251]; Yale Cancer Center Pilot Award; Bureau of Statistics of China [2018LD02]; ``Chenguang Program'' supported by Shanghai Education Development Foundation and Shanghai Municipal Education Commission [18CG42]; Program for Innovative Research Team of Shanghai University of Finance and Economics; and Shanghai Pujiang Program [19PJ1403600].


\newpage
\begin{figure}[h!]
\centering
\subfigure{
\includegraphics[scale=0.35]{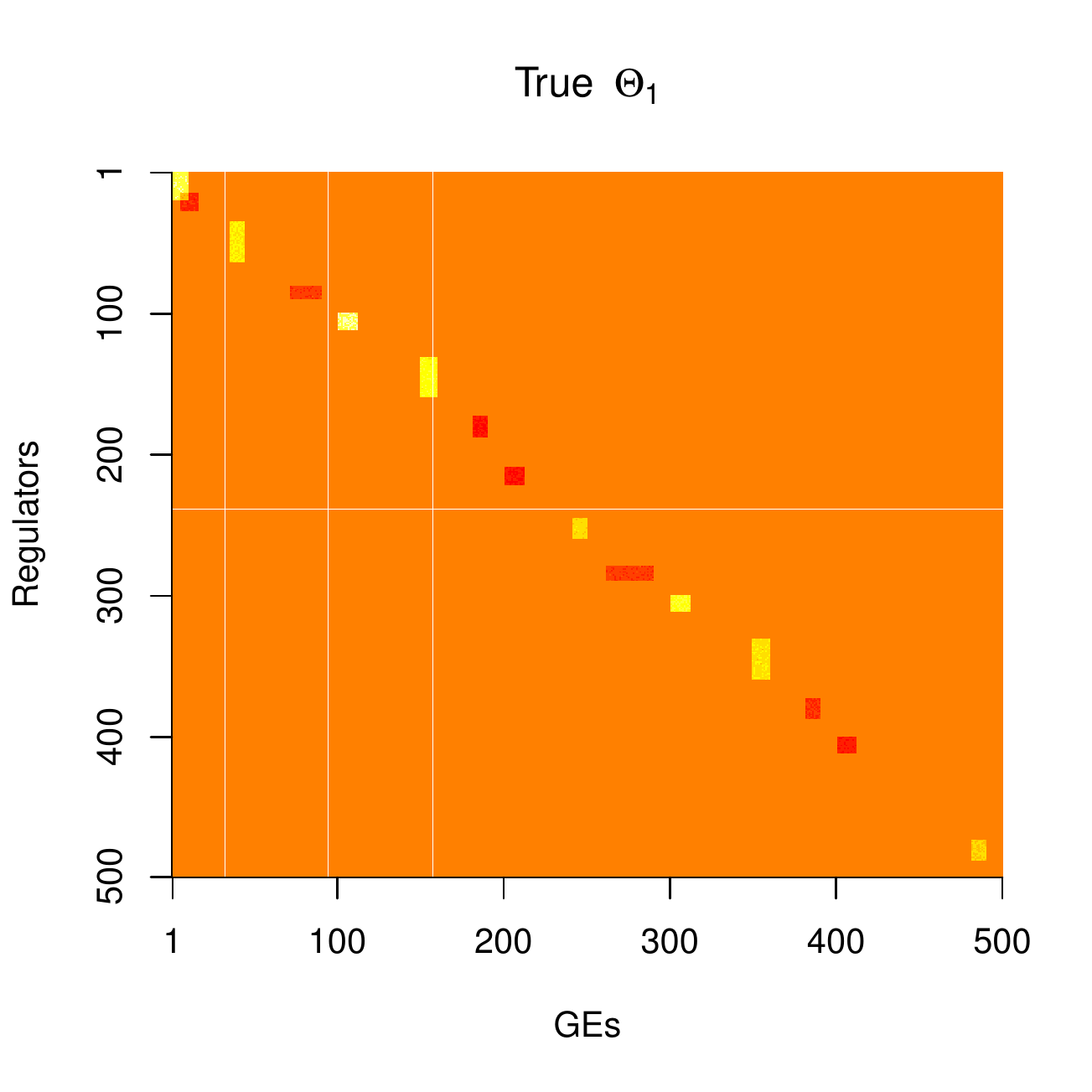}}
\subfigure{\includegraphics[scale=0.35]{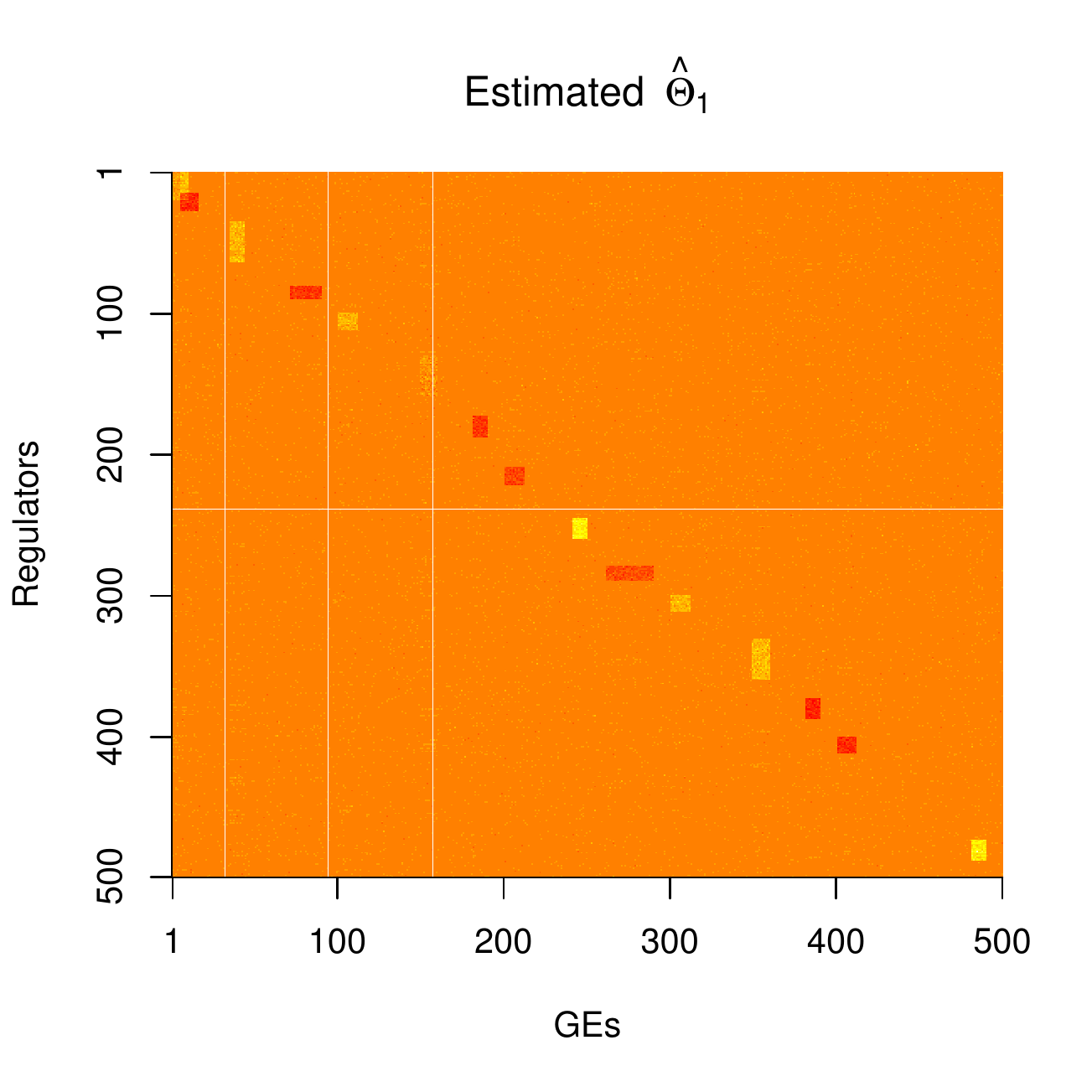}}
\subfigure{%
\includegraphics[scale=0.35]{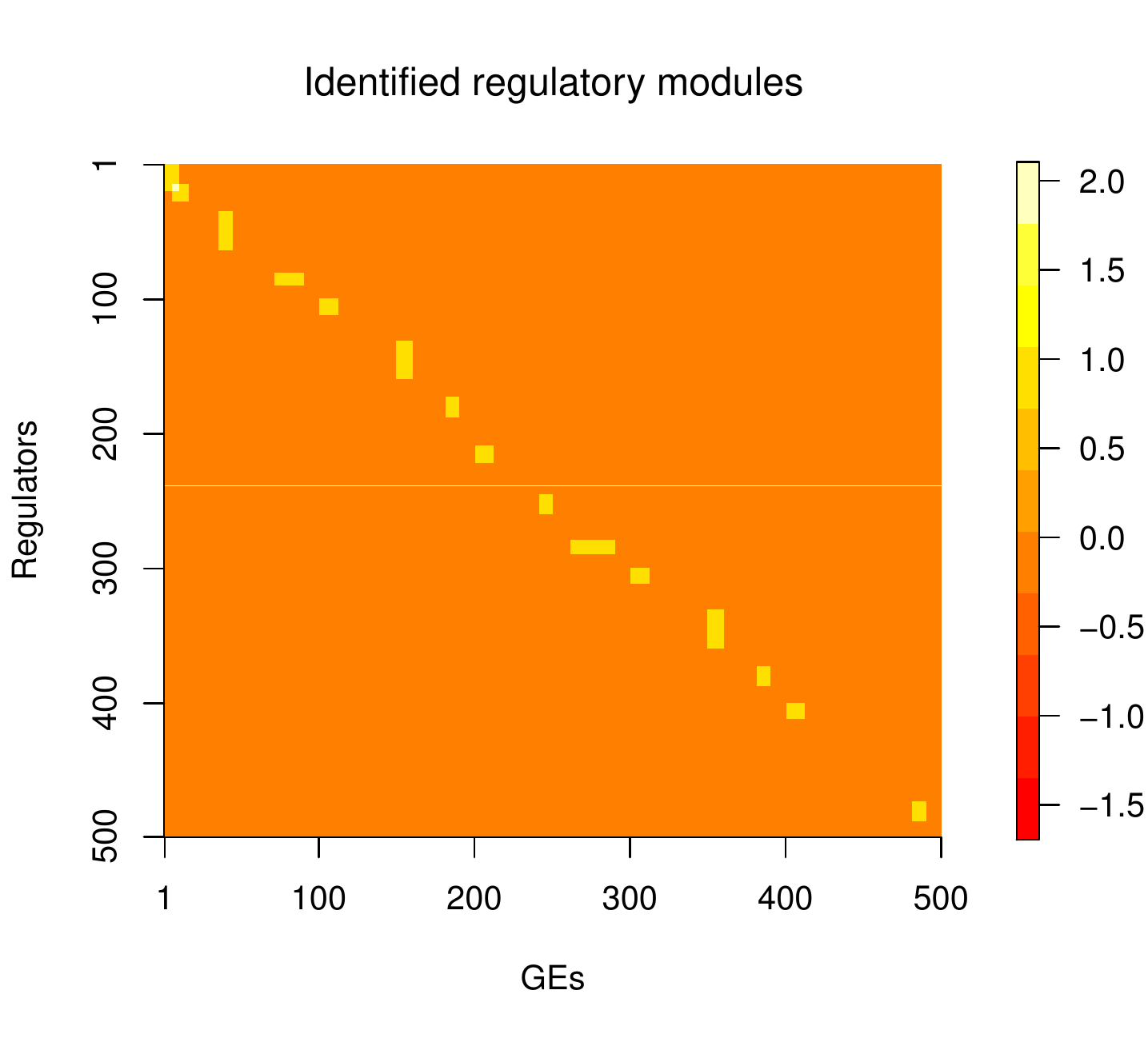}}
\caption{Simulation. Left: true values of regulation under setting $\boldsymbol{\Theta}_1$ and R1; Middle: estimated values; Right: identified regulatory modules.}
\label{fig:theta}%
\end{figure}

\newpage
\begin{table}[h!]
\centering
\caption{Summary results for simulation under setting P1 with a total of 100 true positives: mean (sd) from 200 replicates.}{} \label{tab:t1}
\renewcommand{\tabcolsep}{0.7pc} 
\begin{tabular}{ccl rrrrr} \hline
\multicolumn{3}{c}{}& \multicolumn{2}{c}{$\Theta_1$}&&\multicolumn{2}{c}{$\Theta_2$} \\\cline{4-5} \cline{7-8}
\multicolumn{2}{c}{}& \multicolumn{1}{l}{Approach}&\multicolumn{1}{c}{TP}&\multicolumn{1}{c}{FP}&&
\multicolumn{1}{c}{TP}&\multicolumn{1}{c}{FP}\\	\hline
B1	&	R1	&	Proposed	&	95.94(4.63)	&	11.39(13.83)	&	&	80.06(5.37)	&	6.31(6.02)	\\
	&		&	Alt.1	&	71.90(35.35)	&	20.45(24.20)	&	&	27.06(19.83)	&	1.94(1.12)	\\
	&		&	Alt.2	&	65.20(28.39)	&	28.75(14.03)	&	&	31.69(15.05)	&	7.94(18.32)	\\
&&	Alt.3	&	23.15(3.47)	&	7.45(3.90)	&&	20.35(9.10)	&	19.85(8.43)	\\
	&		&	Alt.4	&	16.80(2.09)	&	122.20(38.35)	&	&	29.85(5.73)	&	127.40(40.31)	\\
	\\
	&	R2	&	Proposed	&	97.30(1.75)	&	5.70(10.99)	&	&	80.72(4.64)	&	20.89(25.90)	\\
	&		&	Alt.1	&	86.60(30.13)	&	13.00(16.06)	&	&	47.00(16.82)	&	6.11(4.92)	\\
	&		&	Alt.2	&	85.15(16.11)	&	39.95(6.87)	&	&	33.61(11.44)	&	19.06(21.78)	\\
	&&	Alt.3	&	23.35(4.18)	&	7.65(2.89)	&&	21.05(5.77)	&	25.79(6.27)	\\
	&		&	Alt.4&	16.95(2.86)	&	126.05(46.47)	&	&	16.00(9.56)	&	71.15(65.52)	\\
	\\
	&	R3	&	Proposed	&	99.70(0.57)	&	8.50(14.60)	&	&	79.40(3.22)	&	36.13(38.10)	\\
	&		&	Alt.1	&	83.68(27.69)	&	12.26(18.29)	&	&	51.14(25.72)	&	8.71(11.69)	\\
	&		&	Alt.2	&	95.90(8.09)	&	54.75(12.48)	&	&	30.50(14.39)	&	4.50(7.60)	\\
	&&	Alt.3	&	27.30(1.63)	&	14.70(17.41)	&&	20.21(7.79)	&	22.11(8.46)	\\
	&		&	Alt.4	&	20.75(2.65)	&	136.65(37.06)	&	&	20.00(8.55)	&	103.05(63.77)	\\	\hline
B2	&	R1	&	Proposed	&	99.80(0.41)	&	14.25(18.95)	&	&	83.90(4.43)	&	12.60(10.56)	\\
	&		&	Alt.1	&	99.80(0.41)	&	57.80(22.75)	&	&	32.00(23.43)	&	14.35(13.92)	\\
	&		&	Alt.2	&	85.80(14.06)	&	55.55(27.20)	&	&	34.75(13.98)	&	18.25(9.48)	\\
	&&	Alt.3	&	27.17(2.46)	&	5.28(1.02)	&	&27.15(8.67)	&	35.15(11.45)	\\
	&		&	Alt.4	&	21.45(2.98)	&	142.05(31.31)	&	&	30.30(10.98)	&	110.10(66.72)	\\
	\\
	&	R2	&	Proposed	&	99.82(0.39)	&	4.12(14.69)	&	&	77.88(3.67)	&	20.81(12.93)	\\
	&		&	Alt.1	&	90.80 (27.98)	&	38.65(21.69)	&	&	42.69(22.46)	&	12.06(8.73)	\\
	&		&	Alt.2	&	77.85(19.63)	&	47.05(16.62)	&	&	21.75(19.49)	&	19.05(22.55)	\\
	&&	Alt.3	&	27.37(2.29)	&	7.79(3.31)	&	&17.45(5.84)	&	18.95(11.87)	\\
	&		&	Alt.4	&	19.60(2.19)	&	135.35(36.02)	&	&	14.95(9.74)	&	52.40(47.98)	\\
	\\
	&	R3	&	Proposed	&	99.35(0.67)	&	12.05(17.72)	&	&	77.77(2.95)	&	9.85(6.67)	\\
	&		&	Alt.1	&	96.45(13.77)	&	50.45(17.38)	&	&	35.65(19.45)	&	21.95(13.06)	\\
	&		&	Alt.2	&	86.45(12.17)	&	46.45(11.91)	&	&	16.50(16.99)	&	10.00(13.13)	\\
	&&	Alt.3	&	28.88(3.14)	&	7.71(2.52)	&	&14.35(4.89)	&	16.35(7.19)	\\
	&		&	Alt.4&	21.25(2.71)	&	140.65(30.84)	&	&	14.95(10.79)	&	65.15(62.45)	\\
	\hline
\end{tabular}
\end{table}

\newpage
\begin{table}[h!]
\centering
\renewcommand{\tabcolsep}{0.7pc} 
\caption{Summary results for simulation under setting P2 with a total of 70 true positives: mean (sd) from 200 replicates.}{} \label{tab:t2}
\begin{tabular}{ccl rrrrr} \hline
\multicolumn{3}{c}{}& \multicolumn{2}{c}{$\Theta_1$}&&\multicolumn{2}{c}{$\Theta_2$} \\\cline{4-5} \cline{7-8}
\multicolumn{2}{c}{}& \multicolumn{1}{l}{Approach}&\multicolumn{1}{c}{TP}&\multicolumn{1}{c}{FP}&&
\multicolumn{1}{c}{TP}&\multicolumn{1}{c}{FP}\\	\hline
B1	&	R1	&	Proposed	&	68.85(0.88)	&	0.40(0.50)	&	&	67.30(3.26)	&	2.30(5.65)	\\
	&		&	Alt.1	&	63.30(15.69)	&	34.80(23.26)	&	&	33.95(20.68)	&	6.68(10.37)	\\
	&		&	Alt.2	&	65.25(4.27)	&	31.85(8.55)	&	&	52.15(10.98)	&	38.30(36.79)	\\
		&&	Alt.3	&	22.25(4.46)	&	7.85(4.49)	&	&	20.95(4.32)	&	23.85(9.28)	\\
	&		&	Alt.4	&	15.30(2.75)	&	129.45(42.29)	&	&	28.10(4.10)	&	127.65(43.63)	\\
	\\
	&	R3	&	Proposed	&	57.15(18.43)	&	10.50(6.36)	&	&	53.90(12.49)	&	2.65(2.21)	\\
	&		&	Alt.1	&	42.55(28.65)	&	18.00(25.46)	&	&	38.25(15.21)	&	4.40(8.18)	\\
	&		&	Alt.2	&	42.50(21.19)	&	24.05(14.57)	&	&	28.00(17.26)	&	3.70(6.14)	\\
		&&	Alt.3	&	24.50(2.50)	&	10.90(8.42)	&	&	18.00(11.31)	&	23.00(24.04)	\\
	&		&	Alt.4&	14.30(1.95)	&	107.05(30.44)	&	&	16.95(5.31)	&	83.85(51.99)	\\
	\\
	&	R3	&	Proposed	&	67.15(6.71)	&	1.40(3.98)	&	&	51.90(13.63)	&	2.55(2.74)	\\
	&		&	Alt.1	&	61.00(19.74)	&	27.80(17.56)	&	&	34.25(6.54)	&	8.10(14.49)	\\
	&		&	Alt.2	&	65.35(5.05)	&	36.35(11.94)	&	&	44.45(15.43)	&	47.00(43.04)	\\
		&&	Alt.3	&22.75(2.90)	&	9.95(7.49)	&	&	15.75(4.88)	&	18.85(12.33)	\\
	&		&	Alt.4 &	15.40(2.26)	&	117.65(30.91)	&	&	19.20(5.69)	&	105.85(61.27)	\\
	\hline
B2	&	R2	&	Proposed	&	69.75(0.55)	&	1.70(6.67)	&	&	67.05(5.88)	&	10.50(11.00)	\\
	&		&	Alt.1	&	69.75(0.44)	&	32.30(11.68)	&	&	65.40(7.38)	&	28.65(30.47)	\\
	&		&	Alt.2	&	66.40(5.23)	&	38.15(30.67)	&	&	46.00(12.02)	&	2.30(5.25)	\\
	&&	Alt.3	&	25.79(5.54)	&	11.05(17.48)	&	&	20.95(4.32)	&	23.85(9.28)	\\
	&		&	Alt.4&	16.65(2.21)	&	136.00(36.41)	&	&	36.25(4.27)	&	150.50(45.17)	\\
	\\
	&	R2	&	Proposed	&	67.15(11.35)	&	1.15(3.77)	&	&	57.10(11.11)	&	4.55(4.08)	\\
	&		&	Alt.1	&	69.80(0.52)	&	33.70(12.69)	&	&	41.85(15.79)	&	16.35(9.42)	\\
	&		&	Alt.2	&	58.15(9.91)	&	36.55(11.76)	&	&	43.55(15.74)	&	11.50(13.61)	\\
		&&	Alt.3	&	26.85(2.89)	&	6.85(2.13)	&	&	18.00(11.31)	&	23.00(24.04)	\\
	&		&	Alt.4&	17.50(2.65)	&	148.40(45.18)	&	&	18.90(6.21)	&	91.40(53.36)	\\
	\\
	&	R3	&	Proposed	&	69.85(0.37)	&	2.80(7.25)	&	&	56.77(8.12)	&	3.69(6.32)	\\
	&		&	Alt.1	&	68.90(4.46)	&	32.95(9.29)	&	&	39.35(12.57)	&	19.35(18.79)	\\
	&		&	Alt.2	&	66.35(6.67)	&	35.45(7.99)	&	&	50.05(11.98)	&	7.15(14.18)	\\
		&&	Alt.3	&26.70(3.37)	&	7.40(3.42)	&	&	15.75(4.88)	&	18.85(12.33)	\\
	&		&	Alt.4&	16.75(1.97)	&	133.65(32.08)	&	&	16.20(6.05)	&	63.40(43.57)	\\
\hline
\end{tabular}
\end{table}

\begin{table}[h!]
\centering
\renewcommand{\tabcolsep}{0.8pc} 
\caption{Analysis of the LUAD data using the proposed method: identified main effects and interactions.}{} \label{tab:t3}
\begin{tabular}{lll rrrrr} \hline
Group	&	Type	&	Gene	&	Main	&	Age	&	Stage	&	Smoking	&	Gender	\\
\hline
	&		&		&		&	0.010	&	-0.031	&	-0.201	&	-0.067	\\
1	&	GE	&	VIT	&	0.006	&		&		&		&		\\
1	&	GE	&	PRH1	&	0.007	&		&		&		&		\\
1	&	GE	&	NOXRED1	&	0.006	&		&		&		&		\\
1	&	GE	&	RYR3	&	0.007	&		&		&		&		\\
1	&	GE	&	SERPINB11	&	0.007	&		&		&		&		\\
1	&	GE	&	ZNF273	&	0.004	&		&		&		&		\\
1	&	GE	&	WRAP53	&	0.003	&		&		&		&		\\
1	&	GE	&	SNORA7B	&	0.006	&		&		&		&		\\
1	&	GE	&	GUCY2F	&	0.007	&		&		&		&		\\
1	&	GE	&	STATH	&	0.007	&		&		&		&		\\
1	&	GE	&	CACNG6	&	0.007	&		&		&		&		\\
1	&	DM	&	WIPI2	&	-0.005	&		&		&		&		\\
3	&	GE	&	LINC00922	&	-0.059	&		&	0.009	&	0.001	&	0.004	\\
3	&	GE	&	NDP	&	-0.059	&		&	0.008	&	0.001	&	0.004	\\
3	&	GE	&	TNMD	&	-0.055	&		&	0.008	&	0.001	&	0.004	\\
3	&	GE	&	IBSP	&	-0.055	&		&	0.008	&	0.001	&	0.004	\\
3	&	GE	&	PWRN1	&	-0.053	&		&	0.008	&	0.001	&	0.004	\\
3	&	GE	&	CACNG3	&	-0.053	&		&	0.008	&	0.001	&	0.004	\\
3	&	DM	&	MIS18A	&	-0.045	&		&	0.007	&	0.001	&	0.003	\\
3	&	DM	&	RRP1	&	-0.036	&		&	0.005	&	0.001	&	0.002	\\
3	&	DM	&	ZDHHC2	&	-0.044	&		&	0.006	&	0.001	&	0.003	\\
9	&	GE	&	ZXDA	&	-0.014	&		&		&		&		\\
9	&	GE	&	EXOSC8	&	0.022	&		&		&		&		\\
9	&	GE	&	EPSTI1	&	0.020	&		&		&		&		\\
9	&	GE	&	UGT2B4	&	-0.012	&		&		&		&		\\
9	&	CNV	&	SLC22A10	&	0.009	&		&		&		&		\\
9	&	CNV	&	PABPC5	&	0.018	&		&		&		&		\\
9	&	DM	&	ATP8A2	&	0.010	&		&		&		&		\\
9	&	DM	&	DHX32	&	0.013	&		&		&		&		\\
15	&	GE	&	KL	&	-0.016	&		&		&		&		\\
15	&	CNV	&	MAP4K4	&	-0.013	&		&		&		&		\\
15	&	CNV	&	KCMF1	&	-0.016	&		&		&		&		\\
15	&	DM	&	SATB2	&	-0.013	&		&		&		&		\\
\hline \multicolumn{8}{r}{{Continued on the next page}}
\end{tabular}
\end{table}

\setcounter{table}{2}
\begin{table}[h!]
\centering
\renewcommand{\tabcolsep}{0.8pc} 
\caption{Continued from the previous page.}{} \label{tab:t3}
\begin{tabular}{lll rrrrr} \hline
Group	&	Type	&	Gene	&	Main	&	Age	&	Stage	&	Smoking	&	Gender	\\
\hline
16	&	GE	&	HIST1H2AA	&	-0.009	&		&		&		&		\\
16	&	GE	&	KCNIP3	&	-0.008	&		&		&		&		\\
16	&	GE	&	LRRTM3	&	-0.011	&		&		&		&		\\
16	&	GE	&	DCLRE1A	&	-0.012	&		&		&		&		\\
16	&	GE	&	PPP1R3D	&	-0.007	&		&		&		&		\\
16	&	GE	&	NHLRC2	&	-0.009	&		&		&		&		\\
16	&	GE	&	NPAP1	&	-0.010	&		&		&		&		\\
16	&	CNV	&	MAP4K4	&	-0.011	&		&		&		&		\\
20	&	GE	&	FTSJ1	&	0.013	&		&		&		&		\\
20	&	GE	&	DGUOK	&	0.012	&		&		&		&		\\
20	&	GE	&	SESN3	&	-0.008	&		&		&		&		\\
20	&	GE	&	CAPZB	&	0.009	&		&		&		&		\\
20	&	CNV	&	PABPC5	&	0.005	&		&		&		&		\\
20	&	CNV	&	MRGPRD	&	0.008	&		&		&		&		\\
20	&	DM	&	IL17D	&	0.008	&		&		&		&		\\
20	&	DM	&	ATP8A2	&	0.006	&		&		&		&		\\
20	&	DM	&	DHX32	&	0.004	&		&		&		&		\\
21	&	GE	&	AFF3	&	-0.106	&		&		&	0.147	&	-0.013	\\
27	&	GE	&	SGPP2	&	-0.009	&		&		&		&		\\
47	&	GE	&	FNIP2	&	-0.027	&		&		&		&		\\
50	&	GE	&	C11orf65	&	0.005	&		&		&		&		\\
68	&	GE	&	DRD3	&	0.012	&		&		&		&		\\
102	&	GE	&	DPRX	&	0.026	&		&		&		&		\\
124	&	GE	&	PRIMA1	&	-0.016	&		&		&		&		\\
178	&	GE	&	FAM217B	&	-0.013	&		&		&		&		\\
304	&	CNV	&	AK4	&	0.014	&		&		&		&		\\
319	&	CNV	&	MIR582	&	-0.024	&		&		&		&		\\
423	&	DM	&	HOXA1	&	-0.027	&		&		&		&		\\
520	&	DM	&	SDE2	&	0.014	&		&		&		&		\\
\hline
\end{tabular}
\end{table}

\begin{table}[h!]
\centering
\renewcommand{\tabcolsep}{0.7pc} 
\caption{Analysis of the SKCM data using the proposed method: identified main effects and interactions.}{} \label{tab:t6}
\begin{tabular}{lll rrrrr} \hline
Group	&	Type	&	Gene	&	Main	&	Age	&	Stage	&	Gender	&	Clark	\\
\hline
	&		&		&		&	-0.176	&	-0.099	&	0.150	&	-0.042	\\
14	&	GE	&	MYCNOS	&	0.003	&	  	&	  	&	  	&	  	\\
14	&	GE	&	MRGPRX3	&	0.004	&	  	&	  	&	  	&	  	\\
14	&	GE	&	MFSD6L	&	0.005	&	  	&	  	&	  	&	  	\\
14	&	GE	&	IMP3	&	0.005	&	  	&	  	&	  	&	  	\\
14	&	GE	&	TBC1D7	&	0.003	&	  	&	  	&	  	&	  	\\
14	&   GE	&	A2M	&	0.004	&	  	&	  	&	  	&	  	\\
14	&	GE	&	NEURL2	&	0.005	&	  	&	  	&	  	&	  	\\
14	&	GE	&	IL24	&	0.004	&	  	&	  	&	  	&	  	\\
14	&	DM	&	MAU2	&	0.004	&	  	&	  	&	  	&	  	\\
14	&	DM	&	ZDHHC4	&	0.004	&	  	&	  	&	  	&	  	\\
14	&	DM	&	ENOX1	&	0.002	&	  	&	  	&	  	&	  	\\
14	&	DM	&	PTPN12	&	0.005	&	  	&	  	&	  	&	  	\\
14	&	DM	&	BRF2	&	0.002	&	  	&	  	&	  	&	  	\\
14	&	DM	&	SYT6	&	0.002	&	  	&	  	&	  	&	  	\\
70	&	GE	&	DSTYK	&	-0.054	&	0.123	&	-0.164	&	  	&	  	\\
71	&	GE	&	GLDN	&	0.044	&	-0.058	&	0.012	&	  	&	  	\\
82	&	GE	&	RBP2	&	-0.029	&	-0.026	&	  	&	  	&	  	\\
124	&	GE	&	SATB2	&	-0.057	&	-0.032	&	0.034	&	  	&	  	\\
153	&	GE	&	RPL36AL	&	0.003	&	  	&	  	&	  	&	  	\\
204	&	GE	&	RNPS1	&	-0.057	&	0.112	&	  	&	0.084	&	  	\\
214	&	GE	&	ARL6IP1	&	-0.014	&	  	&	  	&	  	&	  	\\
573	&	DM	&	DPY19L3	&	0.006	&	  	&	  	&	  	&	  	\\
640	&	DM	&	RABEP1	&	-0.080	&	  	&	-0.071	&	0.010	&	  	\\
647	&	DM	&	SLU7	&	-0.004	&	  	&	  	&	  	&	  	\\
654	&	DM	&	KLHL31	&	-0.023	&	  	&	  	&	  	&	  	\\
696	&	DM	&	GLMP	&	-0.016	&	  	&	  	&	  	&	  	\\
714	&	DM	&	BNIP1	&	-0.025	&	  	&	  	&	  	&	  	\\
759	&	DM	&	MS4A15	&	0.055	&	-0.045	&	  	&	  	&	  \\\hline
\end{tabular}
\end{table}

\clearpage
\setcounter{table}{0}
\renewcommand\thetable{{A\arabic{table}}}
\setcounter{section}{0}
\renewcommand\thesection{{A\arabic{section}}}

\setcounter{equation}{0}
\renewcommand\theequation{{A\arabic{equation}}}

\setcounter{figure}{0}
\renewcommand\thefigure{{A\arabic{figure}}}

\clearpage
\section*{Appendix}
\subsection*{Analysis flowchart }

\begin{figure}[!htb]\centering
\includegraphics[scale=0.8]{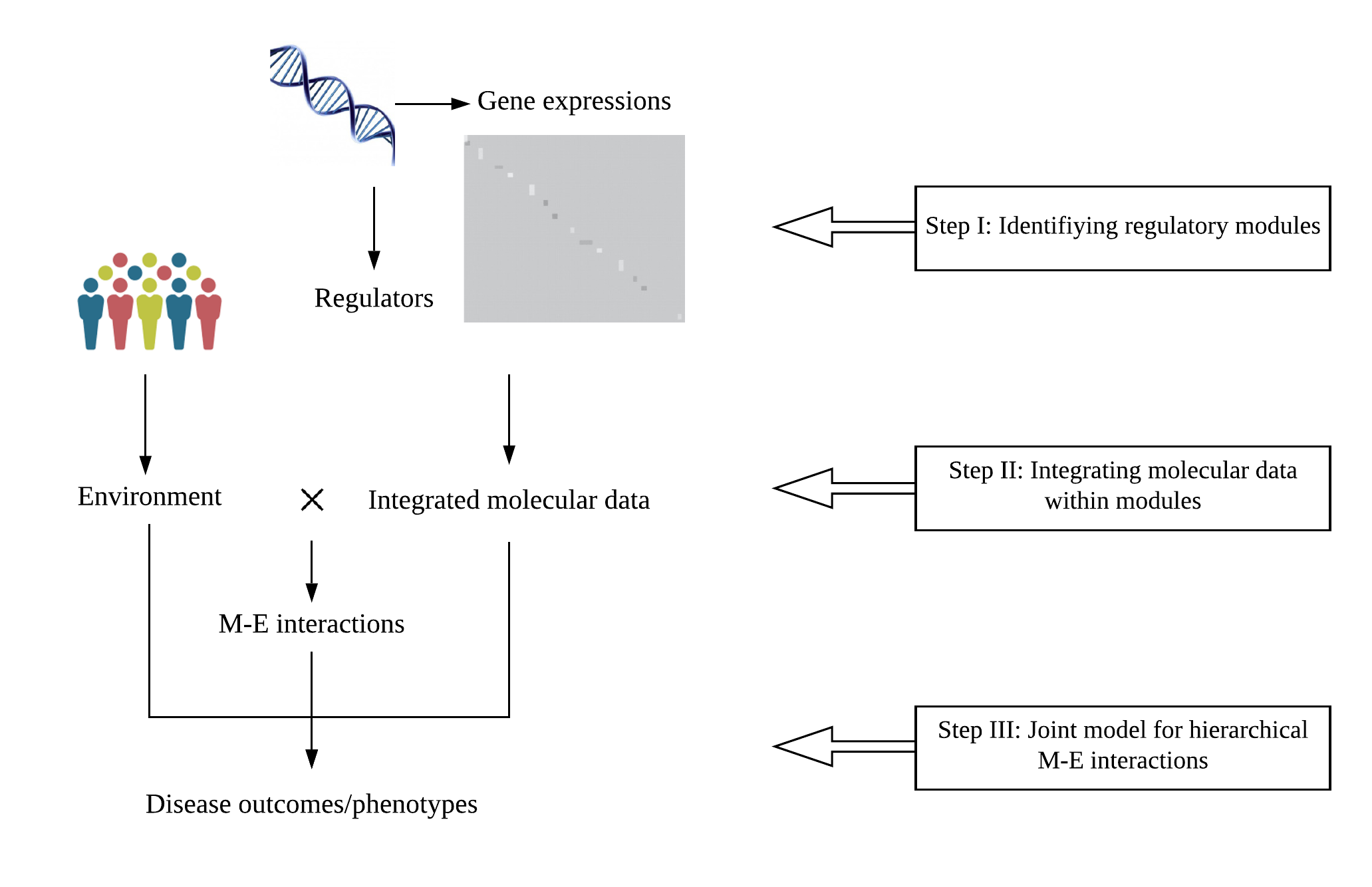}\caption{Flowchart of the proposed M-E interaction analysis.}
\end{figure}

\clearpage
\subsection*{Computational algorithms}

\begin{algorithm}[!h]\caption{\textbf{Identifying regulatory module}}
 \begin{enumerate}
\item Estimate $\boldsymbol{\hat\Theta}$ with objective function (\ref{eq:1}) using R package \texttt{glmnet}.
\item Initialize $s=0$ $\boldsymbol{U}^{(s)}$ as the normalized matrix of $\boldsymbol{\hat{\Theta}}$, where $\boldsymbol{U}^{(s)}$ denotes the remaining regulation relationships at iteration $s$.
\item $s=s+1$. Apply the sparse 2-means clustering to $\boldsymbol{U}^{(s)}$ based on objective function (\ref{eq:2}), and obtain two clusters $\mathcal{C}_s$ and $\bar{\mathcal{C}}_s$ for regulators as well as the weight vector $\boldsymbol{w}_s$ for gene expressions, using R package \texttt{sparcl}.
\item Fix $\mathcal{C}_s$ and $\bar{\mathcal{C}}_s$, and permute the rows of $\boldsymbol{U}^{(s)}$ to calculate weight $w_{s,j}^{*}=b_j/\sqrt{\sum_{j'}b_{j'}^2}$ with $b_j=\left(\frac{1}{q} \sum_{l=1}^q\sum_{l'=1}^qd_{l,l',j}-\frac{1}{q_1}\sum_{l, l' \in \mathcal{C}_s} d_{l,l',j}-\frac{1}{q_2}\sum_{l, l' \in \bar{\mathcal{C}}_s} d_{l,l',j}\right)$, under the null hypothesis of no clusters.
\item Repeat Step 4 $B$ times, and then compute $w_{s,(j)}^{0}=\sum_{k=1}^B w_{s,(j),k}^{*}/B$ with $\boldsymbol{w}_{s,(j),k}^{*}$ being the $j$th order statistic of the weights at iteration $k$ of Step 4.
\item Conduct a two-sample Kolmogorov-Smirnov test to compare $\boldsymbol{w}_s$ and $\boldsymbol{w}_{s}^{0}$.
\item If the test at Step 6 rejects the null hypothesis at significance level 0.05, then $j^{\ast}$ gene expressions with the largest weights are selected, where $j^{\ast}=\arg\max_j \left(w_{(p-j+1)}-w_{(p-j+1)}^{0}\right)-\left(w_{(p-j)}-w_{(p-j)}^{0}\right)$. Denote the corresponding index set as $\mathcal{D}_s$. Update $\boldsymbol{U}^{(s+1)}$ by excluding the information of the identified module $\left\{\mathcal{C}_s,\mathcal{D}_s\right\}$ as
 \begin{equation}
  U^{(s+1)}_{lj}=
    \begin{cases}
      U^{(s)}_{lj}-(\bar{U}^{(s)}_{\mathcal{C}_s, j}-\bar{U}^{(s)}_{\bar{\mathcal{C}}_s, j}), & \text{if}\ l\in \mathcal{C}_s\textrm{ and } j\in \mathcal{D}_s  \\\notag
      U^{(s)}_{lj}, & \text{otherwise},
    \end{cases}
  \end{equation}
  where $\bar{U}^{(s)}_{\mathcal{C}_s, j}=\frac{1}{q_1}\sum_{i\in \mathcal{C}_s} U^{(s)}_{ij}$ and $\bar{U}^{(s)}_{\bar{\mathcal{C}}_s, j}=\frac{1}{q_2}\sum_{i\in \bar{\mathcal{C}}_s} U^{(s)}_{ij}$.
 \item Repeat Steps 3-7 until the test at Step 6 fails to reject the null hypothesis, and return the final regulatory modules $\left\{\mathcal{C}_1,\mathcal{D}_1\right\},\cdots, \left\{\mathcal{C}_S,\mathcal{D}_S\right\}$ with $S+1$ being the termination iteration.
 \end{enumerate}
 \end{algorithm}

\begin{algorithm}[!h]\caption{\textbf{M-E interaction analysis with integrated molecular data}}
 \begin{enumerate}
\item Initialize $t=0$, $\boldsymbol{\Phi}^{(0)}=\boldsymbol{0}$, and $\boldsymbol{res}^{(0)}=\boldsymbol{Y}$, where $\boldsymbol{\Phi}^{(t)}$ and $\boldsymbol{res}^{(t)}$ denote the estimates of $\boldsymbol{\Phi}$ and residual $\boldsymbol{res}$ at iteration $t$.
\item Update $t=t+1$. Optimize $Q({\boldsymbol{\Phi}})$ by cycling through $\boldsymbol{\alpha}$, $\boldsymbol{\beta}_s$, $\boldsymbol{\gamma}$, $\boldsymbol{\eta}_{sm}$ and $\boldsymbol{\tau}_{m}$.
 \begin{enumerate}
 \item Update $\boldsymbol{\alpha}$ with the least squared solution. Let $\tilde{\boldsymbol{Y}}=\boldsymbol{res}^{(t-1)}+\boldsymbol{E}\boldsymbol{\alpha}^{(t-1)}$, then $\boldsymbol{\alpha}^{(t)}=(\boldsymbol{E}'\boldsymbol{E})^{-1}\boldsymbol{E}'\tilde{ \boldsymbol{Y}}$. Update $\boldsymbol{res}^{(t-1)}=\tilde{\boldsymbol{Y}}-\boldsymbol{E}\boldsymbol{\alpha}^{(t)}$.
\item For $s=1, \dots, S$, update $\boldsymbol{\beta}_s$ sequentially. Let $\tilde{\boldsymbol{Y}}=\boldsymbol{res}^{(t-1)}+\boldsymbol{X}_s\boldsymbol{\beta}^{(t-1)}_s+\sum_{m=1}^M(\boldsymbol{E}'_m \odot\boldsymbol{X}'_s)' \left(\boldsymbol{\beta}_s^{(t-1)}\ast\boldsymbol{\eta}_{sm}^{(t-1)}\right)$ and $\tilde{\boldsymbol{W}}_s=(\tilde W_{s1}, \dots, \tilde W_{s, p_s})=\boldsymbol{X}_s+\sum_m (\boldsymbol{E}'_m \odot\boldsymbol{X}'_s)' \odot \left(\boldsymbol{\eta}_{sm}^{(t-1)}\right)'$. Then, if $||\tilde{\boldsymbol{W}}'_s \tilde{\boldsymbol{Y}}||_2<\lambda_1\sqrt{p_s}$, update $\boldsymbol{\beta}^{(t)}_s=\boldsymbol{0}$; Otherwise, update
${\beta}^{(t)}_{sj}=\underset{\beta_{sj}}{\operatorname{arg\,min}}\frac{1}{2}
 ||\tilde{\boldsymbol{Y}}-\sum_{j'\ne j}\tilde W_{sj'}\hat{{\beta}}^{(t)}_{sj'}-\tilde W_{sj}{\beta}_{sj}||_2^2+\lambda_1\sqrt{p_s}||{\beta}_{sj}||_2$, for $j=1, \dots, p_s$, using the R function \texttt{optimize}. Update $\boldsymbol{res}^{(t-1)}=\tilde{\boldsymbol{Y}}-\tilde{\boldsymbol{W}}_s \boldsymbol{\beta}_s^{(t)}$.
\item For $d=1, \dots, p_z$, update $\gamma_d$ sequentially. Let $\tilde{\boldsymbol{Y}}=\boldsymbol{res}^{(t-1)}+\boldsymbol{Z}_d{\gamma}^{(t-1)}_d+\sum\limits_{m=1}^M(\boldsymbol{E}_m\ast\boldsymbol{Z}_d) \left({\gamma}_d^{(t-1)}{\tau}_{md}^{(t-1)}\right)$ and $\tilde{\boldsymbol{W}}_d=\boldsymbol{Z}_d+\sum_m (\boldsymbol{E}_m\ast\boldsymbol{Z}_d){\tau}_{md}^{(t-1)}$, update $
\gamma^{(t)}_{d}=ST\left((\tilde{\boldsymbol{W}}'_d \tilde{\boldsymbol{W}}_d)^{-1}\tilde{\boldsymbol{W}}'_d \tilde{\boldsymbol{Y}},  (\tilde{\boldsymbol{W}}'_d \tilde{\boldsymbol{W}})^{-1}\lambda_2\right)$,  where $ST(a, b)=\text{sign}(a)(|a|-b)_+$ is the soft-thresholding operator. Update  $\boldsymbol{res}^{(t-1)}=\tilde{\boldsymbol{Y}}-\tilde{\boldsymbol{W}}_d\gamma^{(t)}_{d}$.
\item For $m=1,\dots, M$ and $s\in \left\{s: \boldsymbol{\beta}^{(t)}_s\neq 0, s=1,\cdots, S\right\}$, update $\boldsymbol{\eta}_{sm}$ sequentially. Let $\tilde{\boldsymbol{Y}}=\boldsymbol{res}^{(t-1)}+(\boldsymbol{E}'_m \odot\boldsymbol{X}'_s)' \left(\boldsymbol{\beta}_s^{(t-1)}\ast\boldsymbol{\eta}_{sm}^{(t-1)}\right)$ and $\tilde{\boldsymbol{W}}_{sm}=(\tilde W_{sm,1}, \dots, \tilde W_{sm, p_s})=(\boldsymbol{E}'_m\odot\boldsymbol{X}'_s)'\odot \left(\boldsymbol{\beta}^{(t-1)}_s\right)'$. Then, if $||\tilde{\boldsymbol{W}}'_{ms}\tilde{\boldsymbol{Y}}||<\lambda_1\sqrt{p_s}$, update $\boldsymbol{\eta}^{(t)}_{sm}=\boldsymbol{0}$; otherwise, update
$\eta^{(t)}_{sm,j}=\underset{\eta_{sm,j}}{\operatorname{arg\,min}}\frac{1}{2}||\tilde{\boldsymbol{Y}}-\sum_{j'\ne j}\tilde{\boldsymbol{W}}_{sm,j'}{\eta}_{sm,j'}^{(t)}-\tilde{\boldsymbol{W}}_{sm,j}{\eta}_{sm,j}||_2^2+\lambda_1\sqrt{p_s}||\eta_{msj}||_2$ for $j=1, \dots, p_s$. Update $\boldsymbol{res}^{(t-1)}=\tilde{\boldsymbol{Y}}-\tilde{\boldsymbol{W}}_{sm} \boldsymbol{\eta}_{sm}^{(t)}$.
\item For $m=1,\dots, M$ and $d\in \left\{d: \gamma_d^{(t)} \neq 0, d=1, \dots, p_z\right\}$, update $\hat{\tau}_{md}$ sequentially. Let $\tilde{\boldsymbol{Y}}=\boldsymbol{res}^{(t-1)}+(\boldsymbol{E}_m\ast\boldsymbol{Z}_d) \left({\gamma}_d^{(t-1)}{\tau}_{md}^{(t-1)}\right)$ and $\tilde{\boldsymbol{W}}_{md}=(\boldsymbol{E}'_m\odot \boldsymbol{Z}_d')'{\gamma}^{(t-1)}_d$, then ${\tau}^{(t)}_{md}=ST\left((\tilde{\boldsymbol{W}}'_{md}\tilde{\boldsymbol{W}}_{md})^{-1}\tilde{\boldsymbol{W}}'_{md}\tilde{\boldsymbol{Y}}, (\tilde{\boldsymbol{W}}'_{md}\tilde{\boldsymbol{W}}_{md})^{-1}\lambda_2\right).$ Update  $\boldsymbol{res}^{(t)}=\tilde{\boldsymbol{Y}}-\tilde{\boldsymbol{W}}_{md}\tau^{(t)}_{md}$.
\end{enumerate}
\item Repeat Step 2 until convergence. In our numerical study, convergence is concluded if  $\frac{|Q(\hat{\bm\Phi}^{(t-1)})-Q(\hat{\bm\Phi}^{(t)})|}{|Q(\hat{\bm\Phi}^{(t-1)})|}<10^{-4}$.
\end{enumerate}
 \end{algorithm}

\clearpage
\subsection*{Detailed simulation settings}

In Step (e) of simulation, the important main molecular effects and M-E interactions are set as follows.
\begin{itemize}
\item P1 with a total of 100 important effects under the regulation pattern $\boldsymbol{\Theta}_1$: The important main molecular effects consist of $15$ gene expressions and $20$ regulators, among which 30 are involved in one regulatory module and the remaining five are molecular units with individual effects. There are $25$ interactions with gene expressions and $40$ interactions with regulators, relating to one regulatory module and five individual molecular units.
\item P1 with a total of 100 important effects under the regulation pattern $\boldsymbol{\Theta}_2$: The important main molecular effects consist of $25$ gene expressions and $21$ regulators, among which 33 are involved in two regulatory modules and the remaining 13 are molecular units with individual effects. There are $36$ interactions with gene expressions and $18$ interactions with regulators, relating to one regulatory module and nine individual molecular units.
\item P2 with a total of 70 important effects under the regulation pattern $\boldsymbol{\Theta}_1$: The important main molecular effects consist of $15$ gene expressions and $20$ regulators, among which 30 are involved in one regulatory module and the remaining five are molecular units with individual effects. There are $15$ interactions with gene expressions and $20$ interactions with regulators, relating to one regulatory module and five individual molecular units.
\item P2 with a total of 70 important effects under the regulation pattern $\boldsymbol{\Theta}_2$: The important main molecular effects consist of $17$ gene expressions and $21$ regulators, among which 33 are involved in two regulatory modules and the remaining five are molecular units with individual effects. There are $20$ interactions with gene expressions and $12$ interactions with regulators, relating to one regulatory module and four individual molecular units.
\end{itemize}

\clearpage
\subsection*{Additional data analysis results}

\begin{figure}[!htb]\centering
\subfigure{%
\includegraphics[scale=0.5]{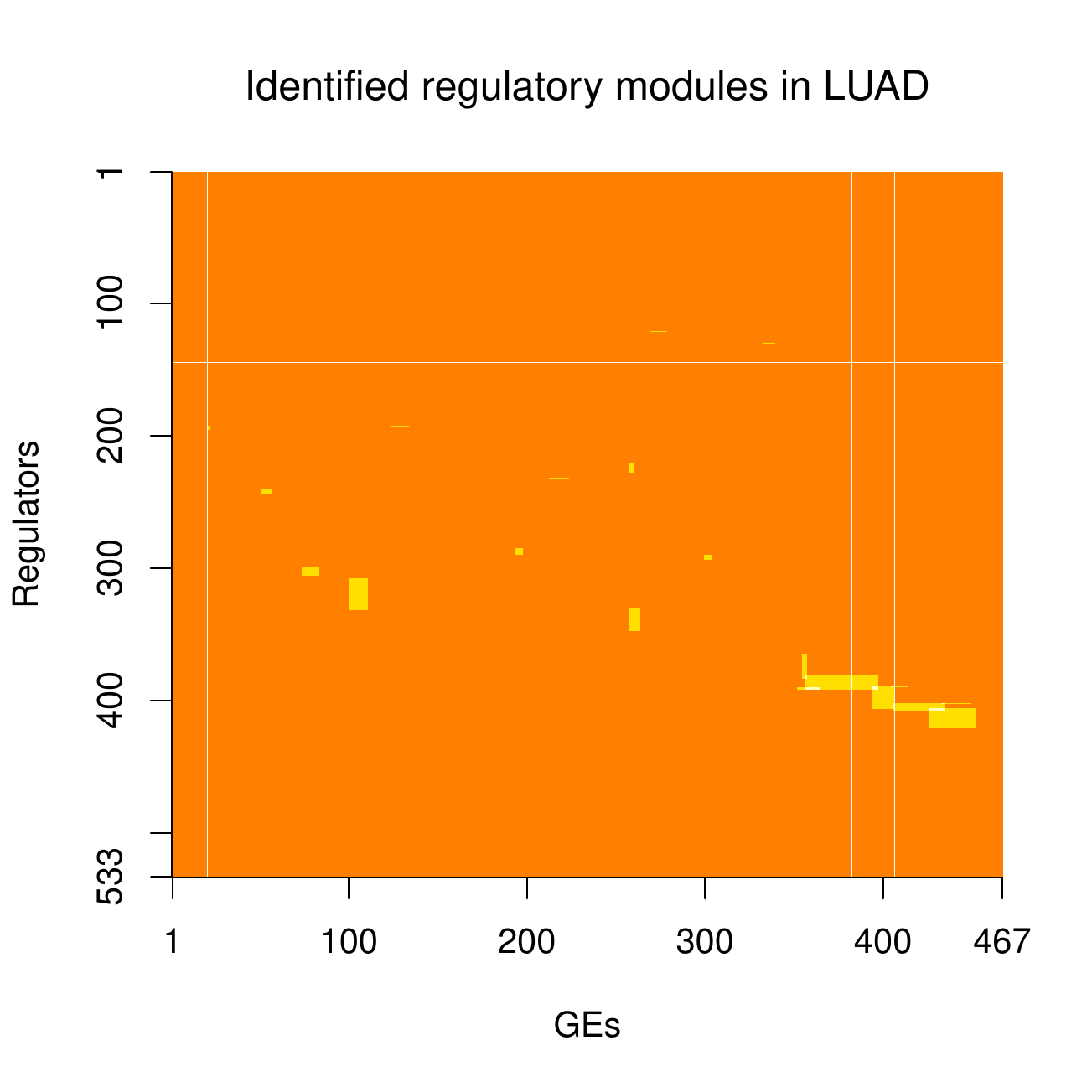}}
\subfigure{\includegraphics[scale=0.5]{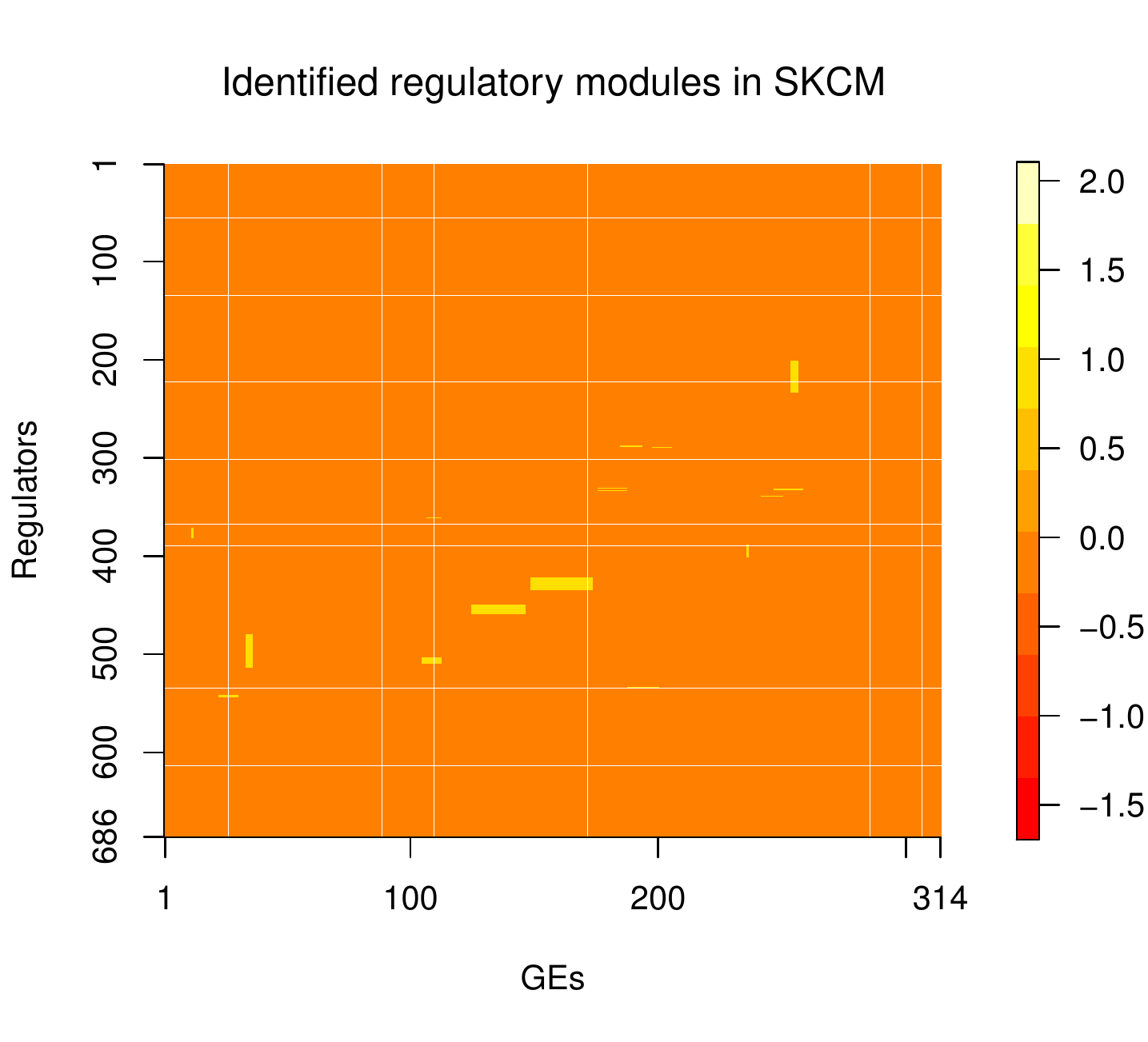}}
\caption{Data analysis: identified regulatory modules.}
\end{figure}

\begin{table}[h!]
\centering
\renewcommand{\tabcolsep}{1pc} 
\caption{Data analysis: numbers of overlapping main molecular effects and M-E interactions (RV-coefficients) identified by different methods.}{} \label{tab:t5}
\begin{tabular}{ll rrrrr} \hline
LUAD & &Proposed	&Alt.1	&Alt.2	&	Alt.3	&	Alt.4	\\
\cline{2-7}
&Proposed	&	62(1)	&	7(0.32)	&	57(0.65)	&	4(0.24)	&	9(0.3)	\\
&Alt.1	&		&	90(1)	&	3(0.31)	&	2(0.08)	&	33(0.58)	\\
Main effects&Alt.2	&		&		&	140(1)	&	6(0.24)	&	11(0.36)	\\
&Alt.3	&		&		&		&	11(1)	&	2(0.08)	\\
&Alt.4	&		&		&		&		&	66(1)	\\
\cline{2-7}
& &Proposed	&Alt.1	&Alt.2	&	Alt.3	&	Alt.4	\\
\cline{2-7}											
&Proposed	&	35(1)	&	0(0)	&	4(0.10)	&	1(0.11)	&	3(0.08)	\\
&Alt.1	&		&	8(1)	&	1(0.03)	&	2(0.28)	&	1(0.12)	\\
Interactions&Alt.2	&		&		&	30(1)	&	1(0.1)	&	1(0.07)	\\
&Alt.3	&		&		&		&	11(1)	&	6(0.27)	\\
&Alt.4	&		&		&		&		&	122(1)	\\
\hline
SKCM &	&	Proposed	&	Alt.1	&	Alt.2	&	Alt.3	&	Alt.4	\\
\cline{2-7}
&Proposed	&	28(1)	&	9(0.47)	&	18(0.62)	&	9(0.52)	&	4(0.39)	\\
&Alt.1	&		&	22(1)	&	7(0.39)	&	12(0.72)	&	6(0.53)	\\
Main effects&Alt.2	&		&		&	35(1)	&	7(0.39)	&	4(0.37)	\\
&Alt.3	&		&		&		&	13(1)	&	4(0.51)	\\
&Alt.4	&		&		&		&		&	10(1)	\\
\cline{2-7}
&	&	Proposed	&	Alt.1	&	Alt.2	&	Alt.3	&	Alt.4	\\
\cline{2-7}
&Proposed	&	12(1)	&	2(0.36)	&	0(0.00)	&	2(0.32)	&	1(0.11)	\\
&Alt.1	&		&	4(1)	&	0(0.01)	&	3(0.66)	&	1(0.19)	\\
Interactions&Alt.2	&		&		&	2(1)	&	0(0.00)	&	0(0.01)	\\
&Alt.3	&		&		&		&	14(1)	&	1(0.20)	\\
&Alt.4	&		&		&		&		&	8(1)	\\
\hline
\end{tabular}
\end{table}

\end{document}